\documentclass[conference]{IEEEtran}
\IEEEoverridecommandlockouts

\usepackage[T1]{fontenc}
\usepackage[utf8]{inputenc}
\usepackage{cite}
\usepackage{amsmath,amssymb,amsfonts}
\usepackage{algorithmic}
\usepackage{graphicx}
\usepackage{url}
\usepackage{hyperref}
\usepackage{breakurl}
\usepackage{textcomp}
\usepackage{xcolor}
\usepackage{booktabs}
\usepackage{array}
\usepackage{multirow}
\usepackage{tabularx}
\usepackage{rotating}
\usepackage{tikz}
\usetikzlibrary{arrows.meta,shapes.geometric,positioning,fit,backgrounds,shadows.blur}

\def\BibTeX{{\rm B\kern-.05em{\sc i\kern-.025em b}\kern-.08em
    T\kern-.1667em\lower.7ex\hbox{E}\kern-.125emX}}

\begin{document}

\title{Adaptive and AI-Augmented Security Testing: A Systematic Survey
of Program Analysis, Feedback-Driven Testing, and Hybrid
Learning-Based Approaches}

\author{
    \IEEEauthorblockN{Michael Wienczkowski}
    \IEEEauthorblockA{
        Dept.\ of Computer Science \& Engineering\\
        Mississippi State University\\
        Starkville, MS, USA\\
        Email: mhw205@msstate.edu
    }
}

\maketitle

%% ─────────────────────────────────────────────────────────────────────
\begin{abstract}
Modern software systems are increasingly developed within rapid
continuous integration and deployment (CI/CD) pipelines, where
ensuring security prior to release presents significant technical
and organizational challenges. Traditional static and dynamic
analysis tools provide valuable structural and behavioral insights,
yet they often operate in non-adaptive workflows and produce large
volumes of warnings requiring manual triage. Simultaneously,
feedback-driven fuzzing techniques and search-based testing
approaches have demonstrated the power of iterative input
refinement guided by execution signals such as coverage and crashes.
More recently, large language models (LLMs) have shown promise in
automated unit test generation and developer assistance; however,
their outputs frequently lack semantic grounding in program
structure and rarely integrate with established program analysis
frameworks.

This paper presents a systematic survey of adaptive and
AI-augmented security testing research, synthesizing work across
five major domains: (1)~structural program analysis for
vulnerability detection, (2)~DevSecOps and continuous security
testing practices, (3)~feedback-driven fuzzing and search-based
testing, (4)~large language model--based automated test generation,
and (5)~emerging hybrid systems integrating program analysis with
adaptive learning mechanisms. Following a structured review protocol
aligned with established systematic literature review (SLR)
guidelines, we analyze fifty-five peer-reviewed studies from
top-tier venues in software engineering and computer security,
drawing on a systematic search of four major databases yielding
22,088 raw records prior to screening.

Our comparative analysis reveals a persistent disconnect between
rich structural program representations---e.g., abstract syntax
trees (ASTs), control-flow graphs (CFGs), and code property graphs
(CPGs)---and adaptive testing mechanisms driven by feedback signals.
We characterize this disconnect as \emph{structural--adaptive
fragmentation}: a systematic separation between semantically precise
systems and adaptive exploration systems that neither paradigm
individually addresses. Existing approaches either emphasize
structural precision without adaptivity, or iterative feedback
without deep semantic grounding, particularly in multi-language
software ecosystems. Notably, no existing system incorporates
human triage signals---the decisions security engineers make when
dismissing false positives or confirming true vulnerabilities---
as a feedback mechanism for refining structural models or adaptive
testing strategies. Recent work on static-analysis-guided LLM
test generation and mutation-directed test refinement represents
early partial progress toward closing this gap, yet no existing
system simultaneously achieves high structural depth, adaptive
feedback integration, and closed-loop incorporation of security
engineer triage decisions into structural model refinement. We
conclude by identifying five open research challenges and
outlining a unified research agenda for semantically grounded,
feedback-driven, polyglot security testing frameworks.
\end{abstract}


%% ─────────────────────────────────────────────────────────────────────
\section{Introduction}
\label{sec:intro}

The acceleration of software delivery through continuous integration
and continuous deployment (CI/CD) has fundamentally transformed
modern software engineering practice. Organizations that once
shipped software quarterly now deploy to production dozens of times
per day~\cite{hilton2016usage}. While these practices enable rapid
feature delivery, they create structural tension with security
assurance: security testing tools and workflows designed for
slower, more deliberate release cycles struggle to keep pace with
high-velocity pipelines without becoming bottlenecks or being
bypassed entirely.

Traditional security testing techniques---static application
security testing (SAST), dynamic analysis, and manual code
review---were not originally designed for high-velocity deployment
environments. SAST tools generate large volumes of warnings whose
manual triage can consume more developer time than the defects
they detect are worth~\cite{christakis2016developers,aloraini2019empirical}.
Dynamic analysis and fuzzing are resource-intensive and difficult
to schedule within the short feedback windows of CI pipelines.
Manual review does not scale. Consequently, security checks are
often deferred, inconsistently applied, or reduced to lightweight
rule-based scans that miss semantic vulnerability patterns.

Simultaneously, three independent research communities have made
substantial progress on complementary aspects of this problem.
The program analysis community has developed expressive,
scalable frameworks for extracting semantic program structure,
including inter-procedural data-flow analysis, taint
tracking, and graph-based vulnerability querying at
industrial scale~\cite{sadowski2015tricorder,calcagno2015infer,
yamaguchi2014cpg,youn2023codeql}. The fuzzing and
search-based testing community has demonstrated that adaptive,
feedback-guided input generation substantially outperforms
random exploration for vulnerability
discovery~\cite{manes2019art,bohme2017directed,stephens2016driller}.
And the natural language processing community has produced
large language models capable of generating syntactically valid,
contextually relevant code---including test code---at a scale
and speed no human team can
match~\cite{schafer2024llm,yang2023whitebox}.

Despite these parallel advances, a fundamental architectural gap
persists. Static analysis systems excel at extracting structural
and semantic program knowledge but remain weakly adaptive:
analysis rules are defined once and applied repeatedly without
refinement based on downstream testing outcomes.
Feedback-driven testing systems iteratively refine inputs using
runtime signals but treat the program under test as a behavioral
black box, rarely leveraging the rich semantic representations
that program analysis provides. LLM-based approaches introduce
generative power at scale but lack principled integration with
formal analysis artifacts, operating on source code as text
rather than as a structured semantic object.

We term this gap \emph{structural--adaptive fragmentation}: the
systematic separation between systems that model program semantics
with high precision and systems that adaptively explore program
behavior using feedback signals. Neither paradigm individually
provides what the other lacks, and their integration---despite
being both technically feasible and practically valuable---remains
largely unrealized in the surveyed literature.

\textbf{Evidence from the corpus.}
The fragmentation is not an abstract architectural observation; it
is directly legible in the surveyed primary studies, which cluster
into two populations that do not overlap. The first population
achieves high structural depth without adaptivity. Yamaguchi
et al.~\cite{yamaguchi2014cpg} and the Joern
framework~\cite{yamaguchi2014joern} represent rich CPG-based structural
modeling but require statically defined traversals with no
feedback-driven refinement; the authors explicitly note that
traversal patterns must be manually constructed and do not respond
to downstream signals. Tsankov et al.~\cite{tsankov2018securify}
demonstrate deep semantic analysis of smart contracts yet operate
in a single-pass, non-iterative mode by design. Li
et al.~\cite{li2018vuldeepecker} and Chakraborty
et al.~\cite{chakraborty2020devign} both leverage structural
representations---code slices and CPGs respectively---for
vulnerability detection, but both are train-once, apply-once
systems with no adaptive feedback loop. The second population
achieves high adaptivity without structural depth.
Sch\"{a}fer et al.~\cite{schafer2024llm}, Pan
et al.~\cite{pan2025aster}, and Wang
et al.~\cite{wang2024hits} each implement iterative LLM feedback
loops that refine tests across multiple rounds using compilation
results and coverage signals; however, all three pass source text
or method signatures to the LLM rather than CPGs, DFGs, or
semantic graphs. Sch\"{a}fer et al.\ explicitly call out the
integration of static and dynamic program analysis as future work
needed to understand test failures, and identify the development
of hybrid feedback-directed and LLM-based techniques as an open
research problem~\cite{schafer2024llm}---directly naming both
dimensions of the gap this survey characterizes. The closest thing to a counterexample in the corpus is
Yang et al.~\cite{yang2023whitebox}, which combines structural
source analysis with iterative LLM feedback---but in the domain
of compiler testing rather than security testing pipelines,
leaving the security-testing instance of the gap unaddressed.
At the practitioner level, Rajapakse
et al.~\cite{rajapakse2022devsecops} identify the disconnect
between structurally powerful SAST tools and the iterative,
feedback-driven nature of CI/CD pipelines as a primary barrier
to DevSecOps adoption, and Feio et al.~\cite{feio2024devsecops}
confirm empirically that security testing in CI/CD pipelines
remains shallow and trigger-based rather than structurally
grounded. The two populations---and the empty region between
them---are made explicit in Fig.~\ref{fig:taxonomy}.

This paper makes the following contributions:

\begin{itemize}
    \item A structured systematic review of fifty-five peer-reviewed
    studies spanning program analysis, DevSecOps, fuzzing,
    LLM-based testing, and hybrid integration, following
    established SLR protocol, drawing on a four-database
    systematic search of 22,088 raw records.

    \item A comparative analysis framework that characterizes
    each study along six attributes---program representation
    type, adaptivity mechanism, feedback integration, LLM
    usage, multi-language support, and evaluation type---making
    structural--adaptive fragmentation empirically visible across
    the corpus.

    \item A formal characterization of structural--adaptive
    fragmentation along four dimensions, grounded in the
    comparative analysis.

    \item An identification of five open research challenges
    and a unified research agenda for semantically grounded,
    feedback-driven, polyglot security testing.
\end{itemize}

\subsection{Scope and Boundaries}

This survey focuses on automated and adaptive approaches to
\emph{security} testing, with emphasis on techniques whose
primary goal is vulnerability detection, security assurance,
or security-relevant test generation. We include work on
general test generation and fuzzing where it directly informs
the security testing landscape or serves as baseline
methodology. We exclude vulnerability patching, exploit
generation, penetration testing tooling, and hardware security
unless directly relevant to a software testing mechanism under
discussion.

\subsection{Paper Organization}

Section~\ref{sec:methodology} describes the review methodology
and research questions. Section~\ref{sec:landscape} presents
a conceptual taxonomy of the research landscape.
Section~\ref{sec:related} positions this survey against
prior surveys in adjacent domains.
Sections~\ref{sec:program_analysis}
through \ref{sec:hybrid} survey each of the five primary
domains. Section~\ref{sec:rqanswers} synthesizes answers
to the four research questions. Section~\ref{sec:comparative}
presents a cross-study comparative analysis.
Section~\ref{sec:fragmentation}
characterizes structural--adaptive fragmentation and identifies
open challenges. Section~\ref{sec:threats} discusses threats
to validity. Section~\ref{sec:conclusion} concludes.

%% ─────────────────────────────────────────────────────────────────────
\section{Review Methodology}
\label{sec:methodology}

This survey follows a structured review protocol informed by
established systematic literature review guidelines for software
engineering research, following Kitchenham and
Charters~\cite{kitchenham2007guidelines}, adapted
for a domain-spanning survey rather than a narrowly scoped
meta-analysis.

\subsection{Research Questions}

The survey is organized around four research questions that
build cumulatively toward the core problem: security engineers
face persistent false positive noise from tools that apply
static rules without learning from prior scan history, human
triage decisions, or runtime evidence. Each RQ characterizes
one dimension of this problem, and together they motivate
the architectural gap that RQ4 identifies.

\textbf{RQ1:} How are structural program representations
(e.g., ASTs, IRs, code property graphs) used for vulnerability
detection and security assurance, and what are their
demonstrated capabilities and limitations in adaptive,
feedback-driven security testing contexts?

\textbf{RQ2:} How do existing automated security testing
systems incorporate runtime feedback signals, and to what
extent does feedback integration reduce false positive rates
and improve vulnerability discovery effectiveness for
security engineers?

\textbf{RQ3:} What are the empirically demonstrated strengths
and limitations of LLM-based automated test generation in
security contexts, particularly regarding semantic grounding
and human feedback integration?

\textbf{RQ4:} What architectural gaps prevent the integration
of semantically grounded structural analysis, adaptive feedback
mechanisms, and human triage signals into a unified security
testing framework deployable in polyglot CI/CD environments?

\subsection{Search Strategy}

A structured literature search was conducted from January~2004
through February~2026 across four primary databases: IEEE~Xplore,
ACM Digital Library, Scopus, and SpringerLink. Search strings were
constructed by combining domain-specific terms across four
conceptual categories:

\begin{itemize}
    \item \textit{Analysis:} ``static analysis,'' ``program
    analysis,'' ``code property graph,'' ``taint analysis,''
    ``abstract interpretation,'' ``symbolic execution''

    \item \textit{Testing:} ``coverage-guided fuzzing,''
    ``greybox fuzzing,'' ``search-based testing,'' ``automated
    test generation,'' ``dynamic analysis''

    \item \textit{AI/ML:} ``large language model,'' ``neural
    vulnerability detection,'' ``graph neural network,''
    ``machine learning security''

    \item \textit{DevOps:} ``DevSecOps,'' ``continuous security
    testing,'' ``CI/CD security,'' ``shift-left security''
\end{itemize}

Boolean combinations of these terms were applied across all
four databases from January~2004 through February~2026.
Table~\ref{tab:search_raw} reports the per-query record
counts returned by each database. Table~\ref{tab:search_counts}
reports the record counts at each stage of the selection
process.

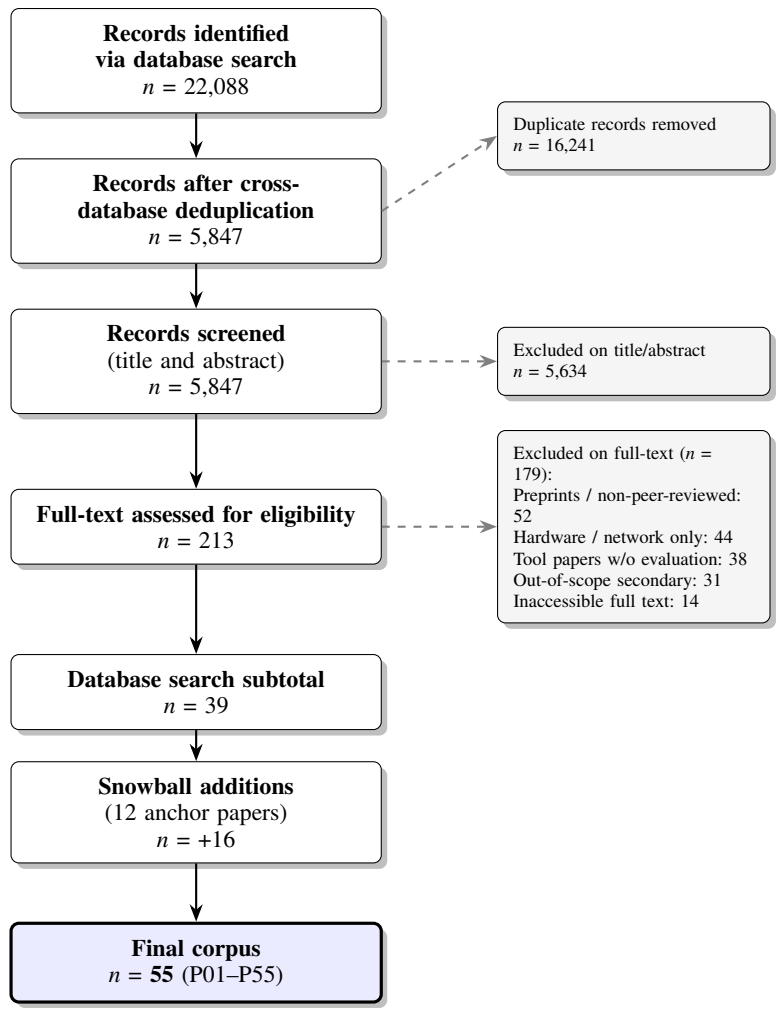
\begin{figure*}[t]
\centering
\begin{tikzpicture}[
    font=\small,
    box/.style={
        draw, rectangle, rounded corners=3pt,
        text width=4.5cm, align=center,
        minimum height=0.9cm, inner sep=6pt,
        fill=white, drop shadow
    },
    exbox/.style={
        draw, rectangle, rounded corners=3pt,
        text width=3.2cm, align=left,
        minimum height=0.9cm, inner sep=6pt,
        fill=gray!8, drop shadow, font=\scriptsize
    },
    arrow/.style={->, thick, >=Stealth},
    exarrow/.style={->, thick, >=Stealth, dashed, gray}
]

%% ── Column x positions ──
\def\cx{0}      % centre column
\def\ex{4.0cm}  % exclusion column (right)

%% ── Main funnel boxes ──
\node[box] (id)   at (\cx,  0)
    {\textbf{Records identified via database search}\\
     \textit{n} = 22,088};

\node[box] (dedup) at (\cx, -2.0)
    {\textbf{Records after cross-database deduplication}\\
     \textit{n} = 5,847};

\node[box] (screen) at (\cx, -4.0)
    {\textbf{Records screened}\\
     (title and abstract)\\
     \textit{n} = 5,847};

\node[box] (ft) at (\cx, -6.2)
    {\textbf{Full-text assessed for eligibility}\\
     \textit{n} = 213};

\node[box] (db) at (\cx, -8.4)
    {\textbf{Database search subtotal}\\
     \textit{n} = 39};

\node[box] (snow) at (\cx, -10.0)
    {\textbf{Snowball additions}\\
     (12 anchor papers)\\
     \textit{n} = +16};

\node[box, fill=blue!8, very thick] (final) at (\cx, -12.0)
    {\textbf{Final corpus}\\
     \textit{n} = \textbf{55} (P01--P55)};

%% ── Main arrows ──
\draw[arrow] (id)     -- (dedup);
\draw[arrow] (dedup)  -- (screen);
\draw[arrow] (screen) -- (ft);
\draw[arrow] (ft)     -- (db);
\draw[arrow] (db)     -- (snow);
\draw[arrow] (snow)   -- (final);

%% ── Exclusion box 1: deduplication ──
\node[exbox, anchor=west] (ex1) at (\ex, -1.0)
    {Duplicate records removed\\
     \textit{n} = 16,241};
\draw[exarrow] (dedup.east) -- (ex1.west);

%% ── Exclusion box 2: title/abstract screening ──
\node[exbox, anchor=west] (ex2) at (\ex, -4.0)
    {Excluded on title/abstract\\
     \textit{n} = 5,634};
\draw[exarrow] (screen.east) -- (ex2.west);

%% ── Exclusion box 3: full-text ──
\node[exbox, anchor=west] (ex3) at (\ex, -6.2)
    {Excluded on full-text (\textit{n} = 179):\\
     Preprints / non-peer-reviewed: 52\\
     Hardware / network only: 44\\
     Tool papers w/o evaluation: 38\\
     Out-of-scope secondary: 31\\
     Inaccessible full text: 14};
\draw[exarrow] (ft.east) -- (ex3.west);

\end{tikzpicture}
\caption{Study selection flowchart following PRISMA-style
reporting guidelines~\cite{kitchenham2007guidelines}.
Starting from 22,088 raw records across four databases,
successive deduplication, screening, and full-text review
stages yielded 39 database studies; backward and forward
snowballing on 12 anchor papers added 16 further studies
for a final corpus of 55 primary studies (P01--P55).}
\label{fig:prisma}
\end{figure*}

The 16 queries yielded 22,088 raw records in total.
After cross-database deduplication---removing records
appearing in more than one database for the same query
concept---5,847 unique records remained for title and
abstract screening. The substantial reduction reflects
the high overlap between IEEE Xplore and Scopus in
particular, which index many of the same IEEE and
Elsevier conference proceedings. Title and abstract
screening against the inclusion and exclusion criteria
reduced the pool to 213 candidates for full-text review.
Full-text assessment excluded 179 records across five
categories: 52 were preprints or non-peer-reviewed
technical reports; 44 addressed hardware, network, or
infrastructure security without code-level software
analysis; 38 were tool announcements or short papers
without accompanying empirical evaluation; 31 were
out-of-scope secondary studies, surveys of surveys, or
position papers; and 14 were inaccessible through
institutional library subscriptions. The database
search thus yielded 39 primary studies. Backward and
forward snowballing on 12 anchor papers added 16 further
studies not returned by any database query, for a final
corpus of 55 primary studies (P01--P55).
Fig.~\ref{fig:prisma} illustrates the complete selection
process as a PRISMA-style flowchart.

\begin{table}[htbp]
\caption{Raw Search Record Counts by Database and Query}
\label{tab:search_raw}
\centering
\small
\setlength{\tabcolsep}{4pt}
\begin{tabular}{lrrrr}
\toprule
\textbf{Query Focus} & \textbf{IEEE} & \textbf{ACM} &
\textbf{Scopus} & \textbf{Springer} \\
\midrule
Q1: Program analysis / & & & & \\
\quad vulnerability detection & 6{,}746 & 764 & 6{,}549 & 886 \\
Q2: Fuzzing / & & & & \\
\quad symbolic execution & 590 & 615 & 1{,}416 & 1{,}835 \\
Q3: LLM / & & & & \\
\quad test generation & 337 & 319 & 663 & 46 \\
Q4: DevSecOps / & & & & \\
\quad CI/CD security & 243 & 20 & 474 & 585 \\
\midrule
\textbf{Database total} & \textbf{7{,}916} & \textbf{1{,}718} &
\textbf{9{,}102} & \textbf{3{,}352} \\
\midrule
\multicolumn{4}{l}{\textbf{Combined raw total}} & \multicolumn{1}{r}{\textbf{22{,}088}} \\
\bottomrule
\end{tabular}
\vspace{4pt}
\hrule
\vspace{4pt}
\parbox{\linewidth}{\footnotesize Search conducted January~2004
through February~2026. Filters applied: peer-reviewed journals
and conference proceedings, Computer Science discipline.
Full query strings are provided in Section~\ref{sec:methodology}.}
\end{table}

\begin{table}[htbp]
\caption{Study Selection Process: Record Counts by Stage}
\label{tab:search_counts}
\centering
\small
\begin{tabular}{lr}
\toprule
\textbf{Selection Stage} & \textbf{Records} \\
\midrule
Raw records across 16 queries / 4 databases   & 22{,}088 \\
After cross-database deduplication            & 5{,}847 \\
After title and abstract screening           & 213 \\
\midrule
\textit{Full-text exclusions:} & \\
\quad Preprints / non-peer-reviewed reports  & $-$52 \\
\quad Hardware / network / infra. security only & $-$44 \\
\quad Tool papers without empirical evaluation & $-$38 \\
\quad Out-of-scope secondary / position papers & $-$31 \\
\quad Inaccessible full text                 & $-$14 \\
\midrule
Database search subtotal                     & 39 \\
Snowball additions (12 anchor papers)        & $+$16 \\
\midrule
\textbf{Final corpus (P01--P55)}             & \textbf{55} \\
\bottomrule
\end{tabular}
\vspace{4pt}
\hrule
\vspace{4pt}
\parbox{\linewidth}{\footnotesize Deduplication estimate based on known overlap
between IEEE Xplore and Scopus indexing of IEEE/Elsevier
proceedings. Exact per-query deduplication counts not
separately trackable without reference management software.}
\end{table}

\subsection{Inclusion and Exclusion Criteria}

Studies were included if they met all three criteria: (a) the
study presented an empirical evaluation, theoretical contribution,
or system design relevant to security testing or program analysis;
(b) the study was published in a peer-reviewed conference
proceedings or journal; and (c) the study contributed structural,
adaptive, or hybrid mechanisms directly relevant to software
security testing or vulnerability detection.

Studies were excluded if they: were available only as preprints
or non-peer-reviewed technical reports; focused exclusively on
hardware security without software applicability; addressed
purely network or infrastructure security without code-level
analysis; or represented tool announcements without accompanying
evaluation.

\textbf{Grey literature policy.}
The exclusion of preprints and non-peer-reviewed material
warrants explicit justification, as relevant work on rapidly
evolving topics such as LLM-based testing may appear first
in grey literature. This survey applies a conservative
inclusion standard: peer review is treated as a necessary
condition for primary corpus membership, because the quality
assessment dimensions defined in Section~\ref{sec:methodology}
(empirical rigor, reproducibility, metric clarity, limitation
transparency) are most reliably present in peer-reviewed
work. Grey literature sources---arXiv preprints, technical
reports, industry white papers, tool documentation, and
practitioner blog posts---were consulted during the search
and snowballing phases to identify emerging directions and
confirm that peer-reviewed corpus coverage was not
systematically lagging in any domain. Two works identified
during this process that fell within the search window but
did not satisfy criterion~(a) are cited in the conclusion
as corroborating evidence for Challenges~3 and~5 respectively,
with their exclusion rationale documented explicitly in
Appendix~\ref{app:matrix}. Two additional works---Li
et al.~\cite{li2025iris} (IRIS, ICLR~2025) and Sheng
et al.~\cite{sheng2025llmvuln} (ACM Computing Surveys,
2025)---were identified via a supplementary grey literature
review and are cited in Section~\ref{sec:threats} and the
conclusion as corroborating evidence for Challenges~1, 3,
and~4, with exclusion rationale documented in
Appendix~\ref{app:matrix}. Standards documents and
practitioner guidance (OWASP, NIST SP~800-115, MITRE ATT\&CK)
are cited where they provide authoritative context for
security testing scope and terminology but are not treated
as primary studies.

\subsection{Snowballing Procedure}

Backward and forward citation snowballing was performed on a
set of 12 high-impact anchor papers identified during initial
search, spanning all five survey domains. Backward snowballing
identified foundational works (e.g., Cousot and
Cousot~\cite{cousot1977abstract}, King~\cite{king1976symbolic},
Weiser~\cite{weiser1981slicing}) whose influence on subsequent
research warranted inclusion despite their age. Forward
snowballing on recent anchor papers (e.g., Sch\"{a}fer et
al.~\cite{schafer2024llm}, Yamaguchi et al.~\cite{yamaguchi2014cpg})
identified emerging integration work not yet well-indexed in
database searches. The combined process yielded a final corpus
of fifty-five primary studies (P01--P55), listed in
Appendix~\ref{app:matrix}.

\subsection{Quality Assessment}

Each primary study was assessed across four quality dimensions:
(1)~\textit{empirical rigor}, including dataset size, benchmark
diversity, and baseline comparisons; (2)~\textit{reproducibility},
including artifact and dataset availability; (3)~\textit{metric
clarity}, including precise definition and consistent reporting
of evaluation metrics; and (4)~\textit{limitation transparency},
including explicit discussion of threats to validity and
generalization boundaries. These assessments informed the
comparative matrix in Section~\ref{sec:comparative} and the
threat analysis in Section~\ref{sec:threats}.

\subsection{Data Extraction Protocol}

For each primary study, the following attributes were extracted:
primary program representation type; adaptivity mechanism (if
any); feedback signal types integrated; use of LLMs or other
generative models; multi-language or polyglot support; and
evaluation methodology and reported metrics. Extraction was
performed by the author across all fifty-five
primary studies. To assess extraction consistency, a second
independent extraction was subsequently performed by the author
on a randomly selected sample of ten studies (18\% of the
corpus, selected via random number generation), covering all
six attribute dimensions. Inter-rater agreement was computed
using Cohen's kappa~($\kappa$) separately for categorical and
ordinal dimensions. Agreement on the three categorical
dimensions (program representation type, LLM usage,
multi-language support) was $\kappa = 0.90$,
interpreted as almost perfect agreement
per Landis and Koch~\cite{landis1977kappa}. Agreement on
the three ordinal dimensions (adaptivity mechanism, feedback
signal integration, evaluation methodology) was
$\kappa = 0.96$. All disagreements were resolved
through re-reading of the primary source until consensus
was reached; no unresolved conflicts remained.

%% ─────────────────────────────────────────────────────────────────────
\section{Research Landscape Overview}
\label{sec:landscape}

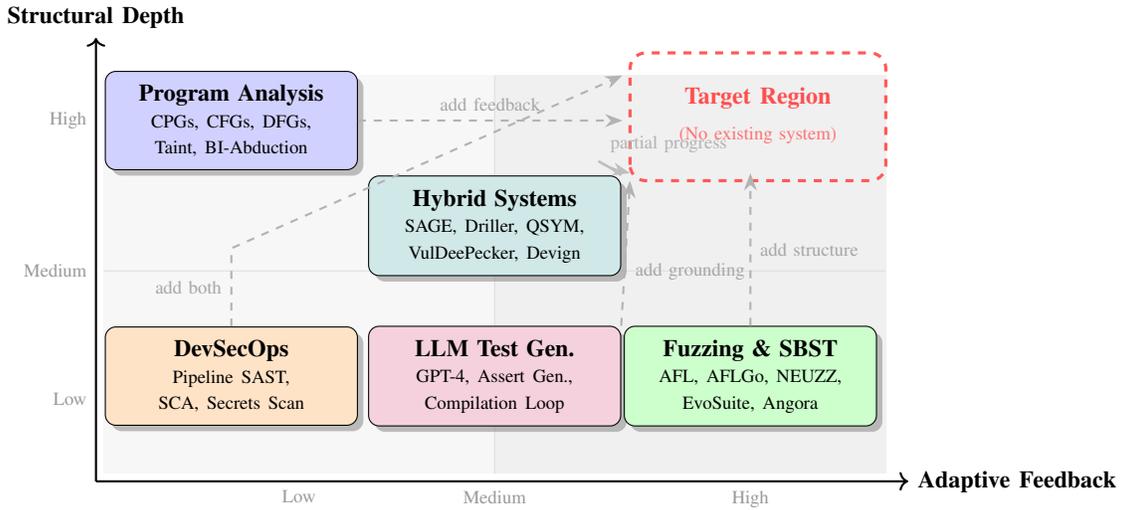
\begin{figure*}[t]
\centering
\begin{tikzpicture}[
    font=\small,
    box/.style={
        draw, rounded corners=4pt, fill=#1,
        text width=3.0cm, align=center,
        minimum height=1.1cm, inner sep=5pt,
        drop shadow
    },
    arrow/.style={->, thick, >=Stealth},
    dasharrow/.style={->, thick, >=Stealth, dashed, color=gray!60}
]

%% --- Background quadrant shading ---
\fill[gray!6]  (0.6,0.5) rectangle (5.8,5.8);   %% low-to-mid adaptive
\fill[gray!12] (5.8,0.5) rectangle (11.0,5.8);  %% high adaptive

%% --- Grid lines (subtle) ---
\draw[gray!25, thin] (5.8,0.5) -- (5.8,5.8);    %% vertical mid
\draw[gray!25, thin] (0.6,3.2) -- (11.0,3.2);   %% horizontal mid

%% --- Axes ---
\draw[thick,->] (0.5,0.4) -- (11.3,0.4)
    node[right] {\textbf{Adaptive Feedback}};
\draw[thick,->] (0.5,0.4) -- (0.5,6.3)
    node[above] {\textbf{Structural Depth}};

%% --- X axis labels ---
\node[below, font=\scriptsize, gray!80] at (3.2,0.4)  {Low};
\node[below, font=\scriptsize, gray!80] at (5.8,0.4)  {Medium};
\node[below, font=\scriptsize, gray!80] at (9.2,0.4)  {High};

%% --- Y axis labels ---
\node[left, font=\scriptsize, gray!80] at (0.5,1.5)  {Low};
\node[left, font=\scriptsize, gray!80] at (0.5,3.2)  {Medium};
\node[left, font=\scriptsize, gray!80] at (0.5,5.2)  {High};

%% -------------------------------------------------------
%% Domain boxes — spaced so no overlap
%% -------------------------------------------------------

%% Program Analysis: high structure, low adaptive  (upper-left)
\node[box=blue!18] (pa) at (2.3,5.2)
    {\textbf{Program Analysis}\\[2pt]
     {\scriptsize CPGs, CFGs, DFGs,\\ Taint, BI-Abduction}};

%% DevSecOps: low structure, low adaptive  (lower-left)
\node[box=orange!22] (dso) at (2.3,1.8)
    {\textbf{DevSecOps}\\[2pt]
     {\scriptsize Pipeline SAST,\\ SCA, Secrets Scan}};

%% LLM Testing: no structure, low-medium adaptive  (bottom-center)
\node[box=purple!18] (llm) at (5.8,1.8)
    {\textbf{LLM Test Gen.}\\[2pt]
     {\scriptsize GPT-4, Assert Gen.,\\ Compilation Loop}};

%% Fuzzing & SBST: low-medium structure, high adaptive  (lower-right)
\node[box=green!20] (fuzz) at (9.2,1.8)
    {\textbf{Fuzzing \& SBST}\\[2pt]
     {\scriptsize AFL, AFLGo, NEUZZ,\\ EvoSuite, Angora}};

%% Hybrid: medium structure, medium adaptive  (center)
\node[box=teal!18] (hyb) at (5.8,3.8)
    {\textbf{Hybrid Systems}\\[2pt]
     {\scriptsize SAGE, Driller, QSYM,\\ VulDeePecker, Devign}};

%% -------------------------------------------------------
%% Target region — upper-right
%% -------------------------------------------------------
\draw[dashed, very thick, red!65, rounded corners=6pt]
    (7.6,4.4) rectangle (11.0,6.1);
\node[font=\small\bfseries, red!70] at (9.3,5.5)
    {Target Region};
\node[font=\scriptsize, red!55] at (9.3,5.0)
    {(No existing system)};

%% -------------------------------------------------------
%% Directional arrows — what each domain must add
%% -------------------------------------------------------

%% Program Analysis → right (needs adaptivity)
\draw[dasharrow] (pa.east)
    -- node[above, font=\scriptsize, color=gray!70]
        {add feedback}
    (7.5,5.2);

%% Fuzzing → up (needs structural grounding)
\draw[dasharrow] (fuzz.north)
    -- node[right, font=\scriptsize, color=gray!70]
        {add structure}
    (9.2,4.5);

%% LLM → true diagonal from top-right corner to target bottom-left
%% Label placed below the line so it clears the Hybrid box
\draw[dasharrow] (llm.north east)
    -- node[below right, font=\scriptsize, color=gray!70]
        {add grounding}
    (7.6,4.4);

%% DevSecOps → diagonal up-right toward target
\draw[dasharrow] (dso.north)
    -- node[left, font=\scriptsize, color=gray!70]
        {add both}
    (2.3,3.5)
    -- (7.5,5.8);

%% Hybrid → curved arc up-right so label clears the box
\draw[arrow, gray!50, thick]
    (hyb.north east)
    .. controls (7.0,4.8) and (7.2,4.6) ..
    node[above right, font=\scriptsize, color=gray!60]
        {partial progress}
    (7.6,4.5);

\end{tikzpicture}
\caption{Conceptual map of the five surveyed research domains
positioned along two axes: \emph{structural depth} (richness
of program semantic modeling) and \emph{adaptive feedback}
(degree to which testing strategies evolve from runtime signals).
The upper-right target region---combining high structural depth
with high adaptive feedback---is occupied by no existing system,
defining the structural--adaptive fragmentation gap this survey
characterizes. Dashed arrows show the direction each domain must
advance; the Hybrid domain represents partial progress but remains
below the target threshold on both axes.}
\label{fig:taxonomy}
\end{figure*}

Before surveying each domain, Fig.~\ref{fig:taxonomy} situates
the five research domains within a unified conceptual map
organized along two axes: \emph{structural depth} (the richness
of program semantic modeling) and \emph{adaptive feedback}
(the degree to which testing strategies evolve based on runtime
signals). Each domain occupies a distinct region of this space,
and the structural--adaptive fragmentation gap is visible as
the empty upper-right quadrant---the region of high structural
depth combined with high adaptive feedback---which no existing
system occupies.

The figure reveals several observations that motivate
the survey organization. Program Analysis (upper
left) and Fuzzing/SBST (lower right) sit at diagonally
opposite corners, representing the two poles of the
fragmentation. DevSecOps (lower left) is operationally
deployed but scores low on both axes, reflecting the
shallow integration documented in Section~\ref{sec:devsecops}.
LLM-based testing sits near the bottom center, capturing
its generative flexibility but lack of formal grounding.
Hybrid systems (center) represent the most advanced current
integration attempts but remain well short of the target
region on both dimensions. The three dashed arrows indicate
the directional research challenges for each primary domain
paradigm; the red dashed box marks the target region that
a unified adaptive framework would need to occupy.

The following sections survey each domain in detail,
organized to build the fragmentation argument cumulatively.

%% ─────────────────────────────────────────────────────────────────────
\section{Related Surveys}
\label{sec:related}

This survey spans territory covered in part by several
prior systematic surveys and literature reviews. Positioning
this work against those surveys clarifies the specific
contribution of the present paper.

\subsection{Fuzzing Surveys}

Man\`{e}s et al.~\cite{manes2019art} provide the most
comprehensive survey of the fuzzing literature, covering
over 60 fuzzing systems organized by input generation
strategy, coverage mechanism, and application domain.
Their taxonomy---black-box, white-box, and grey-box
fuzzing; mutation-based vs.\ generation-based approaches;
coverage-guided vs.\ directed variants---remains the
standard reference framework for the fuzzing community.
The present survey treats Man\`{e}s et al.\ as a primary
study (P23) rather than a related survey because the
fuzzing domain is one of five surveyed here, and because
our analytical focus is on the \emph{integration} of
fuzzing with structural analysis rather than on the
taxonomy of fuzzing techniques themselves.

The key differentiator between the present work and
fuzzing-focused surveys is scope: Man\`{e}s et al.\ do
not address program analysis frameworks, DevSecOps
integration, or LLM-based test generation. They do not
characterize the structural--adaptive fragmentation that
is the central thesis of the present paper, as their scope
is intentionally confined to the fuzzing paradigm.

\subsection{ML-Based Vulnerability Detection Surveys}

Several surveys examine the application of machine learning
to vulnerability detection, covering graph neural network
approaches, sequence models, and representation learning
for code. These surveys characterize the landscape of
learning-based vulnerability detection with greater depth
than the treatment of graph neural network--based detection
in Section~\ref{sec:hybrid} of the present paper.

The present survey differs from this body of work in two
ways. First, it situates learning-based detection within a
broader landscape that includes non-learning structural
analysis, adaptive testing, and generative approaches,
enabling the cross-domain comparative analysis that
single-domain surveys cannot perform. Second, it
specifically examines learning-based vulnerability
detection through the lens of the structural--adaptive
fragmentation thesis---asking not whether machine learning
can detect vulnerabilities (existing surveys establish
that it can), but whether learning-based detection is
coupled to adaptive testing mechanisms that generate,
execute, and refine tests in response to detection
outputs (existing surveys do not address this question,
because its answer is uniformly negative within the
ML-vulnerability-detection domain).

\subsection{DevSecOps and CI/CD Security Surveys}

Rajapakse et al.~\cite{rajapakse2022devsecops} provide a
systematic review of DevSecOps adoption challenges and
solutions that is directly used as a primary study (P15)
in the present survey. Their scope is organizational and
operational: they survey adoption barriers, tooling
integration patterns, and team practices, rather than
the technical architecture of security testing systems.
The present survey extends their work by connecting the
operational constraints they document to the technical
architectural requirements that adaptive security
testing systems must satisfy.

\subsection{LLM for Code Generation Surveys}

The rapid growth of LLM-based code generation has produced
several surveys covering code completion, bug fixing,
test generation, and program synthesis. These surveys
provide broader coverage of the LLM-for-code literature
than Section~\ref{sec:llm} of the present paper.

The present survey focuses narrowly on LLM-based test
generation in security-relevant contexts, with particular
attention to the oracle problem and the absence of
structural program grounding---aspects not emphasized
in code generation surveys whose primary metrics are
functional correctness and developer productivity rather
than security assurance. The present survey is the first,
to the authors' knowledge, to characterize LLM-based
test generation as a dimension of structural--adaptive
fragmentation.

\subsection{Positioning the Present Survey}

The present survey is distinguished from prior work by
three properties that individually exist in prior
surveys but do not co-occur. First, it spans all five
domains simultaneously, enabling the cross-domain
comparative analysis in Section~\ref{sec:comparative}
that single-domain surveys cannot perform. Second, it
organizes the multi-domain synthesis around a single
unifying thesis---structural--adaptive fragmentation---
rather than providing parallel domain summaries without
a cross-cutting argument. Third, it explicitly
identifies the five open research challenges required
to resolve the fragmentation and outlines a research
agenda for unified adaptive security testing, serving
as both a literature review and a research program
specification for the next generation of security
testing systems.

%% ─────────────────────────────────────────────────────────────────────
\section{Structural Program Analysis for Security}
\label{sec:program_analysis}

Structural program analysis forms the semantic foundation of
modern vulnerability detection. By constructing and reasoning
over intermediate representations of program behavior, these
techniques can identify vulnerability classes that neither
dynamic testing nor keyword-based scanning can reliably
detect. This section surveys the primary representation
formalisms used in security analysis (RQ1), the major systems
built upon them, their demonstrated industrial effectiveness,
and the recurring limitations that motivate the adaptive
integration agenda of this survey.

\subsection{Foundational Representations and Theoretical Underpinnings}

The theoretical foundations of structural program analysis
trace to Cousot and Cousot's development of abstract
interpretation~\cite{cousot1977abstract}, which provides a
lattice-theoretic framework for computing sound approximations
of program behavior over all possible inputs and execution
paths. Abstract interpretation allows analysts to reason about
properties such as variable ranges, pointer aliasing, and
information flow without executing the program (at the cost of
precision, since sound over-approximations necessarily admit
false positives).

Weiser's program slicing~\cite{weiser1981slicing} introduced a
complementary decomposition technique: given a variable of
interest at a program point, a slice extracts the subset of
statements that influence that variable's value. Slicing has
proven especially productive in security analysis because many
vulnerability classes---injection attacks, buffer overflows,
use-after-free errors---can be characterized as unsafe
information flows between a source (user-controlled input) and
a sink (security-sensitive operation), making them amenable to
slice-based taint analysis.

Schwartz et al.~\cite{schwartz2010taint} provide a comprehensive
unification of dynamic taint analysis and forward symbolic
execution, clarifying their theoretical relationships and
practical tradeoffs. Dynamic taint analysis instruments program
execution to track the propagation of taint labels from sources
to sinks at runtime, enabling precise detection of taint-style
vulnerabilities with no false positives on the executed path,
but with coverage limited to the inputs exercised. Static taint
analysis operates over all paths simultaneously using
over-approximate data-flow representations, trading precision
for coverage.

Cadar et al.'s EXE system~\cite{cadar2006exe} demonstrated
that symbolic execution could be applied automatically to
generate inputs that trigger memory safety violations,
translating path conditions into constraint satisfaction
problems solvable by off-the-shelf solvers. EXE represented
an early hybrid of structural reasoning and automatic input
generation: a design pattern that would later be central to
hybrid fuzzing systems.

\subsection{Code Property Graphs and Expressive Vulnerability Querying}

A major advance in structural vulnerability analysis was the
introduction of the \emph{code property graph} (CPG) by
Yamaguchi et al.~\cite{yamaguchi2014cpg}. The CPG unifies
three previously separate program representations---the AST,
CFG, and program dependence graph (PDG)---into a single joint
graph structure that can be queried using pattern-matching
languages. By encoding syntactic structure (AST), control flow
(CFG), and data dependence (PDG) in a unified representation,
the CPG enables vulnerability queries that require reasoning
across all three dimensions simultaneously.

For example, a SQL injection vulnerability requires: a
user-controlled value (data source, locatable in the PDG), an
insufficiently sanitized transformation (identifiable in the
AST and PDG), and a database query execution point (control
flow context in the CFG). A CPG query can express all three
conditions conjunctively, yielding precise vulnerability
findings that syntactic or single-representation approaches
miss. Yamaguchi et al. demonstrated CPG-based detection of
previously unknown vulnerabilities in widely deployed C/C++
programs including the PHP interpreter, OpenSSL, and the Linux
kernel, discovering multiple zero-day vulnerabilities confirmed
by vendors.

The CPG was operationalized in the Joern
platform~\cite{yamaguchi2014joern}, which provides an
interactive query environment over CPG representations of C,
C++, Java, JavaScript, Python, and other languages. Joern has
since become a widely-used infrastructure component in both
research and industrial vulnerability analysis workflows,
serving as the basis for CPG-based analyses in subsequent
works~\cite{li2018vuldeepecker,chakraborty2020devign}.

\subsection{Declarative and Polyglot Analysis: CodeQL}

Complementing graph-based approaches, the CodeQL
system~\cite{youn2023codeql} provides a Datalog-based query
language for expressing vulnerability patterns as logical
predicates over program facts extracted from source code. Each
CodeQL analysis begins with database construction, during
which source code is compiled and analyzed to extract a
relational representation of program facts: variable
declarations, call edges, data-flow relationships, and type
information. Analysts then write QL queries that express
vulnerability conditions as joins over these relations.

CodeQL's declarative design offers several advantages for
security analysis. Queries compose modularly, enabling the
construction of complex vulnerability specifications from
simpler predicates. The query language supports inter-procedural
data-flow analysis through a built-in taint tracking library
that handles common propagation patterns across function
boundaries. And CodeQL's multi-language support---covering
C/C++, Java, Python, JavaScript, TypeScript, Go, and
C\#---makes it directly applicable to the polyglot software
ecosystems prevalent in modern CI/CD environments.

Youn et al.~\cite{youn2023codeql} provide a systematic
evaluation of CodeQL across multilingual programs, demonstrating
that cross-language vulnerability queries can be expressed
and evaluated using a unified query framework. Their work
identifies both the expressiveness advantages of declarative
analysis and its current limitations, including incomplete
support for foreign function interfaces (FFIs) and
language-boundary taint propagation, a gap directly relevant
to the polyglot integration challenge identified in this survey.

\subsection{Industrial Deployment: Tricorder and Infer}

The scalability demands of industrial software development
require static analysis systems capable of operating on
codebases with tens of millions of lines of code, integrating
into developer workflows without disrupting productivity, and
maintaining sufficiently low false positive rates to sustain
developer engagement. Two landmark systems document the
challenges and solutions at this scale.

\textbf{Tricorder.} Sadowski et al.~\cite{sadowski2015tricorder}
describe Google's Tricorder program analysis platform, which
integrates multiple static analyzers---including security,
style, and correctness checkers---directly into code review
workflows. Tricorder runs automatically on every proposed code
change, surfacing findings in the code review interface as
suggested fixes co-located with the relevant code. The system
processes hundreds of thousands of code changes per day across
hundreds of distinct analyzers.

A central design insight in Tricorder is the aggressive
management of false positive rates. Google's experience found
that developer trust in automated analysis tools is highly
sensitive to false positive rates: analyzers with rates above
approximately 10\% were routinely dismissed or disabled by
developers~\cite{sadowski2015tricorder}. Tricorder enforces a
requirement that any analyzer surfaced in production maintain
a false positive rate below this threshold, measured against
developer feedback signals (explicit fix actions vs.\
dismissals). This operationalizes the warning prioritization
challenge as a quality gate on analyzer deployment rather than
as a post-hoc filtering problem.

\textbf{Infer.} Calcagno et al.~\cite{calcagno2015infer}
describe Facebook's (now Meta's) Infer static analyzer, which
uses separation logic and bi-abduction to perform
inter-procedural memory safety and resource leak analysis at
scale. Bi-abduction is a technique that automatically infers
function pre- and post-conditions---specifically, the heap
footprints that functions require and modify---enabling
compositional interprocedural analysis without requiring
manually specified contracts. This allows Infer to scale to
large codebases while maintaining precision on memory safety
properties such as null pointer dereferences, use-after-free
errors, and resource leaks.

Calcagno et al.\ report that Infer has been deployed in
Facebook's mobile development pipeline processing millions of
lines of Objective-C, Java, and C code, and has found thousands
of previously unknown bugs in production code. The paper
documents specific deployment challenges including the management
of analysis noise, the integration of incremental analysis to
support per-commit feedback, and the organizational dynamics of
sustaining developer engagement with automated analysis findings
over time.

Critically, both Tricorder and Infer reveal a common limitation
of industrial static analysis at scale: analysis is performed
independently at each commit without learning from previous
analysis runs or incorporating feedback from dynamic testing.
Findings are generated, developers respond (or do not), and
the system resets. There is no mechanism by which runtime
behavior---crashes encountered in testing, coverage data from
CI test suites, sanitizer violations in staging
environments---informs subsequent static analysis priorities.
This architectural pattern is a primary instance of the
structural--adaptive fragmentation identified throughout this
survey.

\subsection{Taint Analysis Frameworks and Security-Specific Tools}

Beyond general-purpose platforms, several analysis systems
target security-specific vulnerability classes through
specialized taint tracking. Livshits and
Lam~\cite{livshits2005java} demonstrated that flow-sensitive
and context-sensitive taint analysis over Java programs, using
a points-to analysis as its foundation, could detect injection
vulnerabilities---SQL injection, cross-site scripting, path
traversal---with substantially higher precision than syntactic
approaches. Their system, BDDBDDB, encoded the analysis as
Datalog queries over binary decision diagram (BDD)
representations of program facts, enabling efficient
computation over large programs.

Gotovchits et al.'s Saluki system~\cite{gotovchits2018saluki}
extends taint-style analysis to binary programs using a
property checking framework over LLVM intermediate
representations. By operating at the IR level rather than
source code, Saluki is applicable to programs whose source
code is unavailable and to compiled components in mixed-source
software systems. The use of LLVM IR as an intermediate
representation also provides a degree of language independence:
any language with an LLVM frontend (C, C++, Rust, Swift, and
others) is analyzable within the same framework.

\subsection{Developer Engagement and Warning Overload}

A persistent challenge in industrial static analysis
deployment is the gap between vulnerability detection
capability and developer remediation behavior. Three studies
in the primary corpus directly examine this gap.

Aloraini et al.~\cite{aloraini2019empirical} conduct a
large-scale empirical study of security warnings generated
by SAST tools across 30 open-source Java projects, analyzing
warning rates, false positive rates, and developer response
patterns over project history. They find that 56\% of SAST
warnings in their corpus were never addressed in project
history, and that warning dismissal rates were higher for
tools with more complex or less actionable finding descriptions.
The study identifies a strong correlation between warning
actionability---the degree to which a warning provides
sufficient context for a developer to understand and fix the
issue---and warning fix rates.

Vassallo et al.~\cite{vassallo2020developers} complement this
with a qualitative study examining how developers engage with
static analysis tools in different organizational and project
contexts. Through interviews and repository analysis, they find
that developer engagement is mediated by organizational factors
including team culture, management support, and tool
integration quality, in addition to technical factors such as
false positive rate and finding precision. Notably, they
observe that developers in projects with active security
ownership---dedicated security engineers or security champions
within development teams---had substantially higher engagement
rates with SAST findings, suggesting that tool effectiveness
is partially a social and organizational question rather than
purely a technical one.

Christakis and Bird~\cite{christakis2016developers} survey
developers at Microsoft about their desired properties of
program analysis tools, finding that developers prioritize
actionability, low noise, and integration with existing
workflows above raw detection power. Developers expressed
willingness to accept some false negatives in exchange for
substantially reduced false positive rates, a preference
that challenges analysis designers who optimize for soundness
or recall. These findings collectively indicate that the
engineering of developer-facing analysis tools requires
treating human factors as first-class design constraints, not
afterthoughts.

\subsection{Structural Analysis Without Adaptivity: The Core Limitation}

Synthesizing across the structural program analysis literature,
a consistent architectural pattern emerges: these systems
achieve remarkable precision and scale in structural semantic
modeling, but they operate as one-shot or episodic analyzers
rather than as adaptive participants in an ongoing testing
process. The following limitations recur across the surveyed
systems:

\textbf{Static rule sets.} Analysis rules are defined by
experts, encoding known vulnerability patterns. Rules do not
evolve based on the analysis history of a particular codebase,
the testing behaviors observed in CI pipelines, or the
vulnerability patterns that previous analysis cycles failed
to detect. This is architecturally analogous to a classifier
that does not update its parameters based on observed
prediction errors.

\textbf{No runtime feedback integration.} Runtime signals
available within modern CI/CD environments---code coverage
from automated test suites, crash reports from staging
environments, sanitizer violations from instrumented
builds---are not consumed by structural analysis systems.
Analysis proceeds from source code alone, ignoring potentially
informative signals about actual program behavior.

\textbf{Independence across analysis cycles.} Each analysis
run is independent of previous runs. The system does not
maintain state about which warnings were previously triaged,
which code regions have been deeply analyzed, or which
vulnerability hypotheses were tested and rejected. This
prevents the kind of iterative refinement that characterizes
effective manual security review.

\textbf{Limited cross-language integration.} While systems
like CodeQL provide multi-language support within a single
query framework, cross-language taint tracking---following
data from a Python web handler through a Java business logic
layer to a C++ database driver---remains largely
unsupported~\cite{youn2023codeql}. This limits applicability
in the polyglot software ecosystems that dominate modern
enterprise and cloud-native development.

These limitations collectively define the first major
dimension of structural--adaptive fragmentation. The rich
semantic representations computed by structural analysis
tools are not currently leveraged to guide adaptive,
iterative security testing. Closing this gap requires
architectures that treat structural analysis outputs not as
terminal findings, but as dynamic inputs to an ongoing
adaptive testing process---a theme developed throughout the
remainder of this survey.

%% ─────────────────────────────────────────────────────────────────────
\section{DevSecOps and Continuous Security Testing}
\label{sec:devsecops}

The DevSecOps paradigm situates security testing within the
operational constraints of modern software delivery. While
Section~\ref{sec:program_analysis} examined the semantic
capabilities and limitations of structural analysis in
isolation, this section examines how security testing
functions---and frequently fails---within CI/CD pipeline
environments. The empirical literature on DevSecOps reveals
a second dimension of structural--adaptive fragmentation:
pipeline automation generates structural analysis artifacts
and runtime telemetry, yet these information streams are
rarely integrated into adaptive testing strategies.

\subsection{The DevSecOps Paradigm and Shift-Left Security}

The term ``DevSecOps'' reflects an organizational and
technical philosophy: security responsibilities and
mechanisms should be integrated into every phase of the
software development lifecycle, rather than concentrated in
a dedicated security review gate at the end of the
development process. The ``shift-left'' principle holds that
security defects found earlier in development are
substantially cheaper to remediate than those found in
staging or production~\cite{rajapakse2022devsecops}.

In practice, shift-left security typically means embedding
automated security checks into CI pipelines that execute on
every code commit. These checks may include: SAST scanning
for known vulnerability patterns; software composition
analysis (SCA) to detect vulnerable dependencies; container
image scanning for known CVEs; secrets detection to prevent
credential leakage; and infrastructure-as-code (IaC) security
linting. When these checks fail, the pipeline blocks the
commit from advancing to deployment---the security quality
gate.

Rajapakse et al.~\cite{rajapakse2022devsecops} conduct a
systematic review of 51 DevSecOps adoption studies, analyzing
reported challenges and solutions across organizational and
technical dimensions. Their synthesis identifies four primary
challenge categories: \textit{cultural resistance}, including
developer friction with security tooling and lack of security
training; \textit{tool integration complexity}, including
incompatibilities between security tools and CI platform
APIs; \textit{false positive management}, including the
developer productivity cost of triaging large volumes of
automated findings; and \textit{performance overhead}, including
the pipeline latency introduced by security scans. Critically,
their analysis finds that technical solutions are frequently
proposed for what are fundamentally organizational and
behavioral challenges---a mismatch that contributes to high
DevSecOps adoption failure rates in their sample.

\subsection{Empirical Studies of Security in CI/CD}

Hilton et al.~\cite{hilton2016usage} provide a large-scale
empirical characterization of CI adoption and usage in
open-source projects, analyzing over 34,000 projects on
GitHub and TravisCI. They find that CI adoption rates have
grown substantially over the study period, with approximately
70\% of active projects using CI by the end of the study
window. However, their analysis of CI configuration files
reveals that security-specific checks are rare: most CI
configurations focus on build, test execution, and code
quality metrics, with explicit security scanning present in
a minority of projects. This empirical baseline suggests
that despite growing DevSecOps discourse, security
integration in CI remains shallow in practice.

Zampetti et al.~\cite{zampetti2020empirical} specifically
characterize the security checks present in CI workflow
configurations across a sample of open-source projects,
analyzing the types, frequencies, and placement of
security-related steps. They find significant heterogeneity
in how projects implement security checks, with most
implementations relying on a single SAST tool or dependency
scanner executed as a post-build step. Integration of multiple
complementary security mechanisms---structural analysis,
dynamic scanning, runtime instrumentation---within a single
pipeline is uncommon. This fragmented integration pattern is
consistent with the tool-centric rather than
architecture-centric view of security in CI/CD that emerges
from the broader literature.

Feio et al.~\cite{feio2024devsecops} provide a more recent
empirical examination specifically focused on continuous
security testing practices, surveying 156 software
professionals about their teams' approaches to automated
security testing in CI/CD contexts. Their findings reveal
that while awareness of DevSecOps practices is high, actual
implementation of continuous adaptive security testing---
where testing strategies evolve based on pipeline feedback---
is rare. The majority of respondents report using static
scanning tools that execute the same rule sets on every
commit without modification. Fewer than 20\% report any
mechanism by which security testing strategies are updated
based on observed testing outcomes. This empirical finding
directly quantifies the adaptive deficit in current DevSecOps
practice.

\subsection{Barriers to SAST Adoption and Effective Integration}

Wadhams et al.~\cite{wadhams2024barriers} examine the specific
barriers that prevent effective SAST tool adoption in
development teams, through a mixed-methods study combining
survey data and developer interviews. Their findings identify
three primary barrier categories that prior literature had
not fully characterized.

\textit{Configuration complexity.} SAST tools require
substantial configuration to reduce false positive rates to
acceptable levels for a given codebase and technology stack.
Default configurations are typically tuned for broad detection
coverage rather than actionability, producing high-noise outputs
that developer teams quickly learn to ignore. The effort
required to configure and maintain custom rule sets is
frequently underestimated during tool adoption, leading to
deployment abandonment after initial pilots.

\textit{Context deficiency.} SAST warnings are often reported
without sufficient context for developers to assess their
exploitability or remediation priority. A warning indicating
a potential SQL injection in a function that processes only
internal administrative data is qualitatively different from
the same warning in a public-facing API endpoint, but most
SAST tools do not incorporate this contextual distinction.
Developers in Wadhams et al.'s study consistently reported
that the absence of runtime context---whether the flagged code
is actually reachable from attacker-controlled inputs in the
deployed application---was the primary driver of dismissal
for security warnings that might otherwise warrant attention.

\textit{Pipeline friction.} Security tools that introduce
noticeable latency into developer feedback loops---the time
from commit to CI result---face resistance regardless of their
detection capability. Developers accustomed to sub-minute
feedback cycles on unit test execution react negatively to
security scans that add five or ten minutes to pipeline
runtime, particularly when those scans produce noisy or
low-actionability results.

These barrier categories collectively suggest that effective
SAST integration in CI/CD is not primarily a detection
algorithm problem, but a human-computer interaction and
systems integration problem: the challenge is not finding
vulnerabilities, but surfacing vulnerability information in a
form that developers can act on efficiently within their
existing workflow context.

\subsection{Security Automation Without Adaptive Feedback Loops}

A consistent observation across the DevSecOps literature is
that pipeline automation operationalizes security checking
without closing a feedback loop. Each CI execution independently
runs the same set of security tools with the same
configurations, generating findings that may or may not be
reviewed, without any mechanism by which the pattern of
previous findings influences the depth, focus, or strategy of
subsequent security checks.

This is architecturally distinct from adaptive testing systems
(Section~\ref{sec:fuzzing}), where feedback signals directly
alter subsequent test generation strategies. In CI/CD security
pipelines, feedback---in the form of warning dismissal rates,
fix latency distributions, and coverage gaps---is generated by
the pipeline but not consumed by it. The pipeline's security
posture does not improve over time based on accumulated
evidence about its own effectiveness.

Rahman et al.~\cite{rahman2019security} study security
practices in infrastructure-as-code artifacts (Ansible and
Chef scripts), identifying recurring security anti-patterns
that persist across long project histories despite the
availability of tools that could detect them. Their finding
that security smells persist over time even in projects with
active development suggests that the absence of adaptive
feedback mechanisms---tools that escalate or re-prioritize
findings that remain unaddressed---contributes to long-term
security debt accumulation.

\subsection{CI/CD as Infrastructure for Adaptive Security}

Despite its current limitations, the CI/CD pipeline
represents an underutilized source of information for
adaptive security testing. Modern CI infrastructure generates
rich telemetry that is currently used primarily for deployment
orchestration rather than security feedback:

\begin{itemize}
    \item \textit{Coverage data} from instrumented test
    execution identifies code regions exercised by existing
    test suites, exposing gaps where vulnerability testing
    has not reached.

    \item \textit{Build failure patterns} identify code
    regions with high churn or instability, which are
    empirically correlated with higher defect
    rates~\cite{bird2009promise}.

    \item \textit{Dependency update events} signal
    third-party library introductions that may bring new
    vulnerability surfaces.

    \item \textit{Performance profiling data} identifies
    hot paths whose optimization pressure may have led to
    security corner cases being deprioritized.
\end{itemize}

None of these signals are currently consumed by structural
analysis or test generation systems in a principled way.
The architecture of DevSecOps pipelines generates exactly
the runtime telemetry that adaptive structural analysis
would require, yet current pipeline tooling does not route
this telemetry back into analysis or generation components.

This constitutes the second major dimension of
structural--adaptive fragmentation: the operational
environment that most clearly motivates adaptive security
testing also represents the most direct source of the
feedback signals that adaptive testing requires, yet the
connection between these two observations is not yet
reflected in production tooling. The research agenda
for closing this gap is developed in
Section~\ref{sec:fragmentation}.

%% ─────────────────────────────────────────────────────────────────────
\section{Feedback-Driven Fuzzing and Search-Based Testing}
\label{sec:fuzzing}

While structural program analysis derives security insight from
examining code as a static artifact, feedback-driven testing
derives it from observing program behavior during execution.
Rather than encoding vulnerability knowledge into fixed rules
applied once, feedback-driven approaches iterate: they generate
inputs, observe what the program does, and use those observations
to generate better inputs on the next cycle. This section surveys
the evolution of coverage-guided fuzzing from its empirical roots
through modern neural and directed variants, examines search-based
testing as a generalization of adaptive input refinement, and
characterizes the core limitation that motivates this survey's
central thesis: adaptivity without semantic grounding.

\subsection{Historical Roots: The Origins of Automated Fuzzing}

The intellectual lineage of modern fuzzing traces to Miller
et al.'s 1990 empirical study~\cite{miller1990reliability},
which demonstrated that purely random input generation---feeding
randomly generated byte streams to UNIX command-line
utilities---caused approximately one third of tested programs
to crash or hang. This finding was both practically significant
and conceptually important: it established that random
exploration, with no knowledge of program structure or semantics,
could reliably expose real vulnerabilities in production software.

Miller et al.'s result established the baseline that the
subsequent three decades of fuzzing research have worked to
improve upon: if uninformed random mutation finds bugs at rate
$r$, what combination of feedback signals and search strategies
can achieve rate $kr$ for maximally large $k$, within practical
resource constraints?

\subsection{Coverage-Guided Greybox Fuzzing}

The most influential answer to this question came with
American Fuzzy Lop (AFL)~\cite{zalewski2013afl}, which
introduced lightweight edge coverage instrumentation as a
feedback signal for input selection. AFL instruments program
branches at compile time with a shared-memory bitmap, where
each entry records whether a particular branch edge (source
block, destination block) has been observed during execution.
After each input execution, AFL checks whether any new bitmap
entries were set---indicating that the input explored a
previously unobserved execution path. Inputs that discover new
edges are retained in a corpus and prioritized for mutation;
inputs that do not are discarded. This simple feedback
mechanism, combined with a suite of byte-level mutation
operators (bit flips, byte substitutions, block deletions,
splicing), proved remarkably effective in practice.

AFL's empirical success catalyzed a generation of research
seeking to understand and improve its behavior analytically.
B\"{o}hme et al.~\cite{bohme2016aflfast} model AFL's input
selection behavior as a Markov chain over the space of
execution paths, showing that AFL's seed selection strategy
implicitly concentrates effort on high-frequency (easy-to-reach)
paths rather than low-frequency (rare or hard-to-reach) paths.
Since vulnerabilities often reside in error-handling or
boundary-condition code that is infrequently exercised under
normal inputs, this implicit bias reduces AFL's effectiveness
precisely where it matters most. AFLFast addresses this by
augmenting the seed selection strategy with an energy
allocation scheme that assigns more mutations to seeds that
exercise rare paths, demonstrating measurable improvements
in crash discovery rates over baseline AFL on standard fuzzing
benchmarks.

FairFuzz~\cite{lemieux2018fairfuzz} extends this insight
by identifying rare branches---branches whose true or false
edges are covered by fewer than a configurable threshold
fraction of corpus inputs---and biasing mutation operators
to preferentially preserve conditions that keep execution
in the neighborhood of those rare branches. By maintaining
branch-covering mutations more aggressively, FairFuzz
increases the rate at which rare paths are exercised, with
empirical improvements in coverage and crash discovery on
both synthetic and real-world targets.

\subsection{Directed Fuzzing and Target-Proximity Metrics}

A parallel line of research addresses the specific challenge
of security testing in contexts where analysts have identified
potentially vulnerable code locations---from static analysis
findings, code review, or patch analysis---and want to
concentrate fuzzing effort toward those locations.

AFLGo~\cite{bohme2017directed} introduces directed greybox
fuzzing, which augments AFL's coverage-based seed selection
with a target-proximity metric. Given a set of target program
locations (specific lines or functions identified as
security-relevant), AFLGo computes an inter-procedural
call graph distance from each program basic block to the
nearest target. During fuzzing, seeds are assigned energy
proportional to a simulated annealing schedule that
progressively shifts priority toward seeds with lower
mean distances to targets, implementing a form of
exploration-exploitation tradeoff: early phases explore
broadly to build diverse corpus coverage; later phases
concentrate energy toward target-proximal inputs.

AFLGo demonstrated effectiveness in directed vulnerability
reproduction (reaching known crash sites more efficiently
than undirected AFL) and in crash triage (reproducing
conditions that trigger specific CVEs faster than
undirected search). It establishes an important precedent:
static structural information---specifically, call graph
distances computed from program structure---can be integrated
into adaptive fuzzing to improve targeting, even when the
integration is one-directional (static analysis informs
fuzzing; fuzzing does not update the static model).

\subsection{Hybrid Fuzzing: Combining Coverage and Constraint Solving}

A fundamental limitation of mutation-based fuzzing is its
inability to generate inputs that satisfy complex conjunctive
path conditions. If reaching a vulnerable code region requires
satisfying a condition such as \texttt{if (crc32(buf) == 0xdeadbeef)},
random byte mutation has negligible probability of generating
a satisfying input by chance, and coverage-guided selection
provides no gradient toward satisfaction. This class of hard
constraints defines an ``exploration barrier'' beyond which
mutation-based fuzzers cannot penetrate without external
assistance.

SAGE~\cite{godefroid2008sage} addressed this at industrial
scale using whitebox fuzzing: SAGE executes a target program
under a symbolic execution engine, collecting path constraints
along the executed path, then negates individual constraints
to generate inputs that reach alternative branches. Applied
to Windows media parsers at Microsoft, SAGE discovered
hundreds of previously unknown security vulnerabilities and
became part of the production security testing pipeline,
representing one of the earliest demonstrations that automated security
testing could operate reliably at industrial scale.

Driller~\cite{stephens2016driller} and
QSYM~\cite{yun2018qsym} refined the hybrid approach for
post-AFL greybox fuzzing contexts. Rather than running
symbolic execution on all paths (as SAGE does), Driller
invokes the Angr symbolic execution
engine~\cite{stephens2016driller} selectively when coverage-
guided mutation stalls, specifically when the fuzzer has not
discovered new coverage in a configurable period. Symbolic execution is
applied only to the stalling seed, using constraint solving
to generate inputs that escape the current coverage plateau.
This selective invocation substantially reduces the overhead
of symbolic execution compared to whole-program analysis
while preserving its ability to cross coverage barriers.

QSYM~\cite{yun2018qsym} addresses performance bottlenecks
in concolic execution by implementing a native binary
analysis engine that avoids the translation overhead of
intermediate representation--based symbolic execution. By
executing natively and collecting constraints incrementally,
QSYM achieves concolic execution performance competitive
with pure fuzzing on many targets while retaining the
constraint-solving capabilities needed to cross hard path
conditions. Yun et al.\ report that QSYM discovers more
unique crashes than AFL, Driller, or standalone angr on
the LAVA-M benchmark suite and on real-world programs
including OpenSSL and libarchive.

The hybrid fuzzing literature establishes an important
empirical result that directly informs this survey's
thesis: structural reasoning (constraint solving over
program path conditions) measurably improves adaptive
testing outcomes when integrated with runtime feedback.
However, the structural knowledge used by hybrid fuzzers
is narrow---path constraints along the current execution
path---rather than the rich semantic representations
(taint flows, vulnerability patterns, data-flow graphs)
that program analysis frameworks compute. This narrow
structural integration represents a partial bridge across
the fragmentation gap, but not a complete one.

\subsection{Neural Approaches to Feedback-Driven Testing}

NEUZZ~\cite{she2019neuzz} introduces a qualitatively
different approach to feedback-driven testing by training
a neural network to approximate the relationship between
input bytes and program branch coverage. Given a seed
corpus, NEUZZ trains a feedforward network that takes
an input buffer as input and predicts which branches will
be covered. Gradient computation over this smooth
approximation identifies input byte positions whose
modification is predicted to most increase branch coverage.
These gradient-guided mutations are applied to existing
seeds, producing new inputs that are then executed and
used to update both the corpus and the model.

NEUZZ's key insight is that while program branch behavior
is not differentiable (branches are discrete), a learned
smooth approximation enables gradient-based optimization
of coverage, effectively bringing continuous optimization
methods to bear on a discrete exploration problem. NEUZZ
demonstrated substantial improvements in edge coverage
and crash discovery compared to AFL on a benchmark
of twelve real-world programs.

The neural approximation approach generalizes the
notion of feedback signal: where AFL uses binary
coverage increments as feedback, NEUZZ uses coverage
predictions as a learned objective. This represents
a step toward learning-based integration between
structural program behavior and adaptive input
generation, though the learned model captures behavioral
rather than semantic program structure.

\subsection{Grammar-Aware and Semantics-Informed Fuzzing}

A limitation shared by mutation-based fuzzers is their
difficulty with structured input formats. Programs that
parse complex formats---XML, JSON, SQL, programming
language source code---reject most randomly mutated
inputs at the parsing stage, before reaching the
security-sensitive logic that deeper execution would
exercise. Superion~\cite{wang2019superion} addresses
this by incorporating a grammar specification of the
target input format into the mutation process. Rather
than mutating raw bytes, Superion parses existing seeds
into abstract syntax trees according to the format
grammar and applies tree-level operators---subtree
replacement, node insertion, node deletion---that
preserve syntactic validity by construction. The result
is that a substantially higher fraction of generated
inputs pass initial format validation and reach deeper
program logic.

Angora~\cite{chen2018angora} takes a different approach
to structured input generation, using dynamic taint
tracking to identify which input bytes influence each
branch condition. Given this byte-to-branch dependency
information, Angora applies gradient descent over the
identified byte positions to satisfy branch conditions
numerically. This \emph{principled search} strategy
avoids the combinatorial challenge of mutating all
input bytes uniformly by focusing effort on the bytes
that actually determine branch outcomes. Angora reported
substantially higher coverage and bug discovery rates
than AFL on the LAVA-M benchmark and on real-world
programs.

Angora's use of taint tracking represents a meaningful
structural integration: program data-flow information
guides the adaptive mutation strategy. However, the
taint information used is dynamic (collected during
execution of specific inputs) rather than static (derived
from whole-program analysis), limiting its ability to
guide exploration of program regions not yet reached by
existing corpus seeds.

\subsection{Search-Based Software Testing}

Search-based software testing (SBST)~\cite{harman2004metrics}
generalizes adaptive input refinement by framing test
generation as a metaheuristic optimization problem. A
fitness function maps test inputs to scalar quality
scores (e.g., branch coverage distance, mutation score,
exception triggering probability), and a search algorithm
(genetic algorithm, simulated annealing, hill climbing)
iteratively generates test inputs that improve fitness.
EvoSuite~\cite{fraser2011evosuite} is the most widely
evaluated SBST system for Java, using a whole-suite
genetic algorithm that evolves entire test suites jointly
rather than individual test cases, improving coverage
diversity and reducing redundancy.

Shamshiri et al.~\cite{shamshiri2015evaluation} evaluate
whether EvoSuite and Randoop~\cite{pacheco2007randoop}---a
feedback-directed random testing tool for Java---generate
tests that find real faults in mature open-source projects.
Their analysis of 1,186 real faults across five projects
finds that automatically generated tests detect only a
small fraction of faults compared to manually written
regression tests. Critically, the faults missed
disproportionately require understanding of semantic
program properties---value constraints, state machine
invariants, security-relevant input combinations---that
coverage-based fitness functions do not capture.

This finding is directly relevant to the security testing
context: coverage is an inadequate proxy for vulnerability
detection. A test suite that achieves high branch coverage
may still fail to exercise the specific input combinations
that trigger a buffer overflow, a type confusion
vulnerability, or an authentication bypass. Security-oriented
fitness functions---taint propagation distances, sanitizer
activation probabilities, authorization boundary violations---
are needed but poorly supported by current SBST frameworks.

\subsection{The Adaptivity--Semantics Tradeoff}

Synthesizing across the feedback-driven testing literature,
a consistent pattern emerges that directly mirrors the
limitation identified for structural analysis in
Section~\ref{sec:program_analysis}: feedback-driven systems
achieve remarkable effectiveness through adaptive exploration,
but their feedback signals are behavioral rather than
semantic. The following specific limitations recur:

\textbf{Behavioral coverage as proxy.} Edge coverage, path
coverage, and crash detection are observable behavioral
signals that correlate with defect discovery---but imperfectly,
and particularly poorly for semantic vulnerability classes.
Vulnerability-specific feedback signals derived from program
structure (taint propagation depth, data-flow constraint
satisfaction, vulnerability pattern proximity) are largely
absent from current adaptive frameworks.

\textbf{Program-as-black-box.} Most greybox fuzzers
instrument programs for coverage but do not extract or
reason over program structure. The fuzzer has no knowledge
of which input bytes correspond to security-sensitive parsing
paths, which branch conditions guard authentication logic,
or which memory operations are taint-reachable from
attacker-controlled inputs. This structural blindness
limits the precision of adaptive prioritization.

\textbf{Shallow static integration.} Where static
information is used (in directed fuzzing for target
proximity, in QSYM for path constraint extraction, in
Angora for taint-guided byte selection) it is used
narrowly and unidirectionally. Static analysis informs
fuzzing as a preprocessing step, but fuzzer outputs
do not update static models. There is no mechanism by
which discovering a crash in one code region informs
re-analysis of structurally adjacent regions for related
vulnerability patterns.

\textbf{Coverage-oracle gap.} Feedback-driven testing
systems detect crashes and sanitizer violations as
security-relevant oracle signals, but these represent
a narrow slice of the security vulnerability space.
Logical vulnerabilities---authentication bypasses,
authorization failures, information disclosure through
timing channels---do not necessarily crash the program
and are therefore invisible to crash-based oracles.

These limitations define the third major dimension of
structural--adaptive fragmentation: adaptivity without
semantic grounding. The complementary strengths of
structural analysis (Section~\ref{sec:program_analysis})
and adaptive testing are apparent---structural analysis
provides exactly the semantic program knowledge that
adaptive testing lacks, while adaptive testing provides
exactly the iterative refinement capability that structural
analysis lacks---yet principled integration of these
strengths into a unified system remains largely unrealized,
as detailed in Section~\ref{sec:hybrid}.

%% ─────────────────────────────────────────────────────────────────────
\section{Large Language Models for Automated Test Generation}
\label{sec:llm}

Large language models represent a qualitatively new force
in automated testing: rather than mechanically applying
predefined transformations to existing inputs or
specifications, LLMs generate test code by generalizing
from patterns observed in large training corpora of human-
written code and tests. This generative capability offers
the prospect of test generation that captures semantic
intent---generating assertions that check meaningful
program properties, not merely structural coverage---at
a speed and scale that no search-based or mutation-based
approach can match. This section surveys the empirical
evidence for LLM-based test generation, examines their
integration into fuzzing pipelines, analyzes their
limitations in security-relevant contexts, and
characterizes the semantic grounding deficit that
motivates the hybrid integration agenda of this survey.

\subsection{Foundations: Neural Models for Code Understanding}

The application of neural language models to code
generation has a history predating the transformer era,
but the introduction of large-scale transformer
architectures pre-trained on code
corpora~\cite{tufano2020assert} substantially shifted
the capability frontier. Pre-trained models can generate
syntactically valid code in multiple programming
languages, produce contextually appropriate identifier
names and comments, and complete partial code fragments
in ways that respect API conventions learned from
training data.

Tufano et al.~\cite{tufano2020assert} apply this
capability specifically to assert statement generation:
given a unit test method body without assertions, a
transformer model trained on test-production code pairs
generates candidate assert statements that check the
expected behavior of the method under test. Their
evaluation demonstrates that transformer-generated
assertions match developer-written assertions substantially
more often than random or template-based baselines,
establishing that neural models can capture testing intent
in ways that purely structural or random approaches cannot.

\subsection{Empirical Evaluation of LLM Test Generation at Scale}

The most comprehensive empirical evaluation of modern
LLM-based test generation is provided by Sch\"{a}fer et
al.~\cite{schafer2024llm}, who evaluate five LLMs (including
GPT-4 and earlier GPT models) on automated unit test
generation for 25 open-source Python and JavaScript
projects. Their experimental design is carefully controlled:
for each function under test, they generate tests using
each LLM with standardized prompts, compile and execute
the results, and measure compilation success rate, test
execution success rate, branch coverage, and---critically---
fault detection capability measured against a mutation
testing framework.

Key quantitative findings from Sch\"{a}fer et al.\ include:

\begin{itemize}
    \item Compilation success rates ranged from 57\% to
    89\% across models and languages, with GPT-4
    performing best and smaller models substantially worse.

    \item After compilation filtering, test execution
    success rates were higher, ranging from 72\% to 94\%,
    indicating that most compilable tests execute without
    errors.

    \item Branch coverage improvements over no-test
    baselines were modest but consistent: generated test
    suites achieved 30--52\% branch coverage on average,
    compared to 0\% without any tests---but substantially
    below the 70--80\% achievable by EvoSuite on the same
    targets.

    \item Mutation scores---the fraction of synthetic
    code faults (mutants) detected by generated tests---
    were substantially lower than coverage metrics
    suggested: many generated tests executed code paths
    but made assertions weak enough that mutants survived
    undetected. This ``weak oracle'' problem was the
    primary limitation identified by the authors.
\end{itemize}

The weak oracle finding is particularly significant for
security contexts. A test that calls a function and asserts
only that it returns without throwing an exception will
pass for both correct and vulnerable implementations,
providing no security assurance. The challenge of
generating tests with strong, semantically meaningful
assertions---assertions that would fail if a security
property were violated---is precisely the oracle problem
that LLM-based generation has not yet solved, and that
structural program knowledge could potentially address.

\subsection{Iterative Prompting and Compilation Feedback}

A practical refinement in LLM-based test generation is
the use of iterative prompting loops that incorporate
execution feedback. When a generated test fails to
compile, the compilation error is appended to the prompt
and the model is re-queried, effectively asking it to
fix its own output. Sch\"{a}fer et al.\ and subsequent
work demonstrate that this simple feedback loop
substantially improves compilation success rates---
in some settings reducing compilation failure rates
by 40--60\% relative to single-shot generation.

This iterative loop represents a rudimentary form of
feedback-driven adaptation: the generation strategy is
modified based on an execution signal (compilation result).
However, the feedback signal used---compilation
success or failure---is syntactic rather than semantic.
The model is corrected when it generates code that does
not compile, but not when it generates code that compiles
and executes but fails to detect security-relevant
behavior. Closing this semantic feedback gap (using
oracle violations, coverage gaps, or structural analysis
findings as generation feedback) is an open research
problem.

\subsection{Property-Based Testing and Stronger Oracles}

Vikram et al.~\cite{vikram2023property} investigate
whether LLMs can generate property-based tests---tests
that specify general invariants (e.g., ``for all valid
inputs, the output satisfies property P'') rather than
specific input-output pairs. Property-based tests offer
stronger security-relevant oracles than example-based
tests: a property such as ``no SQL statement constructed
from user input contains unescaped special characters''
directly encodes an injection-prevention invariant.

Vikram et al.\ find that current LLMs can generate
syntactically valid property-based tests in frameworks
such as Hypothesis (Python) and QuickCheck (Haskell) with
moderate success, but that the generated properties are
frequently trivial (e.g., asserting that a function
returns a value of the correct type) or incorrect (e.g.,
asserting properties that do not actually hold for the
function under test). More complex, security-relevant
properties---those requiring understanding of the
function's role in a larger system, its trust assumptions,
or its invariant boundaries---were rarely generated
correctly without substantial contextual scaffolding.
This finding reinforces the case for structural grounding:
generating meaningful security properties requires
understanding of program semantics that raw source code
text does not fully encode.

Two recent systems make partial progress on the oracle
problem from complementary directions. Antal et
al.~\cite{antal2025vulnwitness} investigate using
GPT-4 specifically to generate \emph{vulnerability-
witnessing} unit tests---tests whose passing or failing
behavior distinguishes vulnerable from patched code
versions. By framing the generation task around a
known vulnerability (providing the CVE description and
the patch diff as context), they demonstrate that
LLM-generated tests can serve as regression oracles for
specific known vulnerabilities, substantially
outperforming coverage-directed test generation on
vulnerability-detection metrics. The key enabling factor
is the provision of vulnerability-specific semantic
context, consistent with the structural grounding
thesis. Harman et al.~\cite{harman2025meta} report
Meta's Automated Compliance Hardening system, which
uses mutation testing scores---rather than coverage---
as the feedback signal driving iterative LLM test
generation. By exposing the LLM to surviving mutants
(program variants that existing tests fail to
distinguish), the system directs generation toward
semantically meaningful distinctions rather than
structural coverage, achieving mutation score
improvements that coverage-guided generation does not
reach. Together, these systems demonstrate that richer
oracle-oriented feedback signals can meaningfully
improve LLM test generation quality, directly
supporting Challenge~3 in Section~\ref{sec:fragmentation}.

\subsection{LLM-Assisted Fuzzing: Grammar Synthesis
and Seed Generation}

Recent work has moved beyond using LLMs as standalone
test generators toward integrating them as components
within adaptive testing pipelines, particularly for
input grammar synthesis and seed generation in fuzzing
contexts.

Yang et al.~\cite{yang2023whitebox} demonstrate that
LLMs can generate semantically diverse compiler test
programs that, when used as seeds for compiler fuzzers,
discover previously unknown compiler bugs in GCC and
LLVM. The key insight is that LLMs can generate programs
that are syntactically and semantically valid (satisfying
the language specification) while exploring unusual
language features or combinations that compiler test
suites generated by hand or by random generation rarely
produce. Yang et al.\ combined LLM-generated seeds with
a differential testing oracle---comparing compiler outputs
across multiple compilers to identify discrepancies
indicative of compilation bugs---and discovered multiple
previously unknown bugs confirmed by compiler maintainers.

Zhang et al.~\cite{zhang2025lowcost} address the challenge
of fuzzing programs that consume non-textual input
formats---binary protocols, specialized file formats,
network packet structures---where conventional mutation
fuzzing is largely ineffective due to format sensitivity.
Rather than manually specifying input grammars (a labor-
intensive process requiring expert knowledge of each
format), Zhang et al.\ use LLMs to synthesize input
generators: programs that produce syntactically valid
instances of the target format. Given a description of
the format (from documentation or example inputs), the
LLM generates a Python program that produces random valid
instances, which are then used as seeds for AFL-style
coverage-guided mutation. Zhang et al.\ demonstrate that
LLM-synthesized generators discover substantially more
unique crashes than mutation from a small set of manually
provided seeds, and at substantially lower engineering
cost than manual grammar specification.

These results establish that LLMs can function as
\emph{structural hypothesis generators} when supplied
with sufficient domain context: they can synthesize
structured inputs that exercise program behavior
inaccessible to unstructured mutation. However, the
structural context supplied in these systems is
informal---natural language format descriptions or
example inputs---rather than formal program analysis
artifacts. Supplying formal structural information
(data-flow paths to vulnerable sinks, taint source
characterizations, vulnerability-relevant AST patterns)
as LLM context remains largely unexplored.

Two recent lines of work represent the closest existing
approximations to formal structural grounding in
LLM-based test generation, and are directly relevant to
Challenge~1 of Section~\ref{sec:fragmentation}.
Roy Chowdhury et al.~\cite{roychowdhury2024static}
demonstrate that providing static analysis
artifacts---call graphs, method signatures, and usage
examples extracted by a program analysis pass---as
structured prompt context substantially improves
LLM-generated unit test quality for Java methods
compared to prompts using source code text alone.
Their system performs a static extraction step before
each generation request, representing the first
systematic demonstration that formal program analysis
artifacts can serve as effective LLM prompt grounding.
Wang et al.~\cite{wang2024hits} (HITS) apply a
complementary slicing-based approach: rather than
enriching the prompt context globally, they partition
the test generation problem by slicing methods at
the statement level and generating targeted tests for
individual slices. This method-slicing strategy
improves branch coverage by ensuring each LLM call
targets a semantically bounded unit, directly
alleviating the context window constraints that arise
when entire codebases are supplied as prompt context. Both systems remain
unidirectional---static analysis informs generation,
but generation outcomes do not update or refine the
underlying structural models---yet they establish the
feasibility of the static-to-LLM coupling direction
that a fully closed adaptive loop would require.

\subsection{Context Window Constraints and Hallucination Risks}

The practical deployment of LLMs for test generation
faces two technical limitations that are particularly
consequential in security contexts.

\textit{Context window constraints.} LLMs process a
finite context window---typically 8K to 128K tokens in
contemporary models---that limits the amount of program
context that can be supplied in a single generation
request. For large, complex programs, this means that
the function under test can be provided, but much of
the codebase that defines its security context---the
callers that supply user-controlled inputs, the downstream
functions that consume its outputs, the authentication
and authorization mechanisms that govern its invocations---
cannot fit within the context window. Tests generated
without this broader context may be syntactically valid
but semantically incomplete, failing to capture the
trust relationships that security-relevant testing
requires. The constraint is further compounded in
multi-language systems: Pan et al.~\cite{pan2025aster}
(ASTER) demonstrate LLM-based unit test generation
across Java, Python, and Kotlin, finding that
cross-language context---the behavior of callers or
callees written in a different language---is effectively
invisible to the generation model, since no single
context window can span an inter-language call boundary
with adequate semantic fidelity. This observation
directly anticipates Challenge~4
(Section~\ref{sec:fragmentation}) and establishes
that multi-language test generation is an open problem
even within the LLM paradigm.

\textit{Hallucination of APIs and behaviors.} LLMs
trained on large code corpora sometimes generate
references to API methods, class names, or constants
that do not exist in the actual codebase under test,
producing tests that fail to compile or that silently
test wrong behavior. In security contexts, hallucinated
API calls may generate tests that appear to exercise
security-relevant code paths but actually call
non-existent mock methods, producing a false sense of
security coverage. This risk is heightened when the
codebase uses domain-specific frameworks or custom
security libraries not well-represented in training data.

Both limitations suggest that effective LLM-based
security test generation requires mechanisms for
supplying precise, relevant structural context---rather
than raw source code or documentation---to the generation
process. Structural program analysis artifacts (CPG
subgraphs, taint path summaries, call graph neighborhoods)
are more compact, semantically precise, and context-
efficient than raw source, making them natural candidates
for structured LLM prompting---a hypothesis central to
the research agenda proposed in
Section~\ref{sec:fragmentation}.

\subsection{Generative Automation Without Semantic Grounding}

The LLM-based test generation literature reveals the
third and most recent dimension of structural--adaptive
fragmentation. LLMs introduce a genuinely new form of
adaptivity---generative flexibility guided by natural
language instructions and learned patterns---that
differs qualitatively from both static rule application
and coverage-guided search. But this adaptivity is
currently decoupled from formal program semantics in
two ways.

First, LLMs are trained on source code as text, not as
structured semantic objects. The model learns statistical
patterns in token sequences, which captures substantial
syntactic and idiomatic knowledge but does not ground
generation in the formal properties---data-flow
relationships, taint propagation paths, vulnerability
invariants---that security-relevant testing requires.

Second, the runtime feedback used in current LLM test
generation loops is compilation and execution success,
not semantic correctness relative to security properties.
A test that compiles, executes, and passes is accepted
regardless of whether its assertions are strong enough
to detect vulnerability-relevant behavior.

These two gaps---lack of structural grounding in
generation and lack of semantic feedback in
refinement---define precisely the integration points
where structural program analysis could most productively
augment LLM-based test generation. This integration
is the focus of the hybrid approach literature reviewed
in Section~\ref{sec:hybrid} and the research agenda in
Section~\ref{sec:fragmentation}.

%% ─────────────────────────────────────────────────────────────────────
\section{Hybrid and Emerging Integration Approaches}
\label{sec:hybrid}

The preceding four sections document a consistent pattern:
structural analysis offers semantic precision without
adaptivity; feedback-driven testing offers adaptivity
without semantic grounding; LLM-based generation offers
generative flexibility without formal grounding; and
DevSecOps pipelines offer operational telemetry without
architectural integration. Hybrid approaches represent
attempts---partial, domain-specific, and often
unidirectional---to bridge these gaps. This section
surveys four classes of hybrid integration, analyzes
what each achieves and where each falls short, and
synthesizes the observations into a taxonomy of
integration patterns that characterizes the field's
current state.

\subsection{Symbolic Execution: The Original Hybrid}

The concept of hybrid security testing predates the term
by decades. Symbolic execution, introduced by
King~\cite{king1976symbolic} and extended by
Cadar and colleagues~\cite{cadar2006exe,cadar2013symbolic},
is itself a hybrid: it combines structural reasoning
(maintaining symbolic representations of path conditions)
with dynamic-style exploration (following execution
paths through the program). Rather than reasoning about
all paths simultaneously through abstract
interpretation, symbolic execution follows one path at
a time, accumulating the conjunction of branch conditions
encountered along that path as a symbolic formula.
When a branch is reached, the path condition is negated
and solved by a constraint solver to generate an input
that takes the alternative branch---enabling systematic
path enumeration through program state space.

Cadar and Sen's retrospective~\cite{cadar2013symbolic}
surveys three decades of symbolic execution research,
documenting both the technique's demonstrated effectiveness
in finding security vulnerabilities and its fundamental
scalability challenges. Path explosion---the exponential
growth of distinct execution paths with program size and
loop depth---limits symbolic execution to programs of
moderate size or to bounded execution depths. Memory
modeling limitations, particularly for programs with
complex heap structures or pointer aliasing, introduce
unsoundness that may cause vulnerability paths to be
missed. And constraint solver performance---the cost of
deciding satisfiability for complex path conditions
involving non-linear arithmetic or bit-vector
operations---creates performance bottlenecks that prevent
exhaustive path coverage of real-world programs.

P\u{a}s\u{a}reanu and Visser~\cite{pasareanu2009survey}
survey innovations addressing these scalability challenges,
including compositional symbolic execution (analyzing
procedures independently and composing summaries),
abstraction-refinement for symbolic state space
reduction, and parallel symbolic execution over
distributed compute resources. Despite these advances,
symbolic execution remains most effective for moderate-
size programs or targeted analysis of specific code
regions, making it a complementary component in hybrid
systems rather than a standalone solution for large-scale
security testing.

\subsection{Industrial Hybrid Fuzzing: SAGE}

SAGE~\cite{godefroid2008sage} represents the first
demonstration that hybrid testing---combining dynamic
execution with symbolic constraint solving---could
operate effectively at industrial scale. Applied to
Windows media parsing libraries at Microsoft, SAGE's
generation-based whitebox fuzzing approach (executing
a seed input symbolically, negating path conditions to
generate branch-diverse inputs, and iterating) discovered
hundreds of previously unknown vulnerabilities over
multiple deployment years.

The industrial scale and impact of SAGE established
several important precedents for hybrid security testing
research. First, it demonstrated that the engineering
challenges of applying symbolic execution to real,
complex programs---handling floating-point constraints,
system calls, and complex memory operations---were
surmountable with sufficient engineering investment.
Second, it established the generation-based hybrid
architecture (execute seed $\to$ collect constraints
$\to$ negate and solve $\to$ generate new input $\to$
repeat) as a viable and effective template. Third, it
demonstrated measurable security impact---specific CVEs
discovered, vulnerability classes eliminated---that
provided a template for subsequent hybrid system
evaluation.

\subsection{Selective Hybrid Fuzzing: Driller and QSYM}

Where SAGE applies symbolic execution comprehensively,
Driller~\cite{stephens2016driller} and QSYM~\cite{yun2018qsym}
apply it selectively, using symbolic execution as a
targeted supplement to coverage-guided mutation fuzzing
rather than as the primary exploration mechanism.

Driller monitors AFL's progress and detects coverage
stagnation: periods during which the fuzzer has not
discovered new edges despite continued mutation. When
stagnation is detected for a given seed, Driller invokes
the Angr symbolic execution engine on that seed,
executing symbolically from the current path state and
using constraint solving to generate inputs that advance
beyond the stalling branch. The result is a hybrid
system that uses AFL's efficient mutation for broad
coverage exploration and symbolic execution for targeted
penetration of coverage barriers.

Driller's selective invocation strategy represents an
important architectural pattern: using feedback signals
(coverage stagnation) to trigger structural reasoning
(symbolic execution). This is closer to a bidirectional
coupling than most hybrid systems: the dynamic feedback
signal (stagnation) triggers the structural analysis
component, rather than structural analysis being applied
only as a fixed preprocessing step. However, the feedback
remains narrow---stagnation detection---and the structural
analysis component (symbolic execution) does not
persistently update a structural model of the program
that would inform subsequent fuzzing cycles.

QSYM's contribution is primarily performance: by
implementing concolic execution natively rather than
through an intermediate representation, QSYM reduces
the overhead of symbolic execution sufficiently to
enable more frequent invocation within hybrid fuzzing
workflows. The performance gains allow QSYM to run
concolic execution more continuously rather than only
on stagnation, achieving higher coverage and bug
discovery rates. Yun et al.\ report that QSYM + AFL
discovers more unique crashes than Driller + AFL on
several benchmark programs, suggesting that performance
improvements in the symbolic execution component
directly translate to improved hybrid effectiveness.

\subsection{Graph Neural Networks for Vulnerability Detection}

A distinct class of hybrid integration applies machine
learning to structural program representations for
vulnerability prediction. Rather than using structural
representations to guide test generation or input mutation,
these systems train classifiers that predict vulnerability
likelihood from program structure, enabling automated
triage and prioritization of code regions for manual
review or targeted testing.

VulDeePecker~\cite{li2018vuldeepecker} operationalizes
program slicing for learning-based detection. For each
API call in a target program that is potentially relevant
to a vulnerability class (e.g., buffer operations for
CWE-119, string functions for format string vulnerabilities),
VulDeePecker extracts the backward and forward data-flow
slices---the set of program statements that influence the
API call's arguments (backward slice) and are influenced
by its return value (forward slice). These ``code gadgets''
are concatenated and encoded as sequences for a bidirectional
LSTM classifier. Evaluated on the NVD/SARD vulnerability
database, VulDeePecker demonstrated a false positive rate
of approximately 21\% and a false negative rate of
approximately 6\% on the evaluated vulnerability classes,
substantially outperforming token-level detection baselines.

VulDeePecker's structural contribution is the use of
data-flow slices as the unit of analysis rather than
raw code tokens or file-level snippets. The slice
encoding captures data dependencies that span multiple
functions and files, encoding semantic relationships
that sequential models over raw code cannot represent.
This demonstrated that the choice of program
representation---not merely the choice of model
architecture---is a first-order determinant of detection
quality.

Devign~\cite{chakraborty2020devign} extends this by
encoding richer structural information through graph
neural networks. Rather than encoding slices as sequences,
Devign constructs a joint program graph that combines
AST edges, CFG edges, data-flow edges, and control-
dependence edges into a single heterogeneous graph for
each function. A gated graph recurrent network processes
this graph to produce a function-level vulnerability
prediction. Evaluated on a large corpus of real-world
C programs with CVE-annotated vulnerabilities from
four large open-source projects (QEMU, FFmpeg, OpenSSL,
libcurl), Devign demonstrates accuracy improvements
of 10--15 percentage points over sequence-based models
on vulnerability detection, confirming that richer
structural representations yield improved prediction
quality.

The learning-based vulnerability detection literature
establishes three empirically grounded findings
directly relevant to this survey's thesis. First,
structural program representations carry vulnerability-
predictive signal beyond what syntactic or textual
representations capture---establishing that investing
in structural analysis is justified by downstream
security outcomes. Second, richer structural
representations (CPG-derived graphs) yield better
detection than simpler ones (slices, token sequences),
suggesting that more complete structural integration
would be valuable. Third, and most importantly for
the fragmentation argument: the output of these
systems is a vulnerability likelihood score, not an
executable test, not an adaptive strategy, and not
a feedback signal. The structural knowledge encoded
by learned models is not connected to any mechanism
for generating tests, guiding exploration, or
updating analysis priorities in response to predicted
vulnerability locations. The integration is
unidirectional and terminal: structure $\to$ prediction.

\subsection{Domain-Specific Integration: Securify}

Securify~\cite{tsankov2018securify} provides a
particularly instructive example of tight structural--
security integration within a constrained domain:
Ethereum smart contract security analysis. Securify
combines datalog-based semantic analysis with a
library of compliance and violation patterns to
provide sound (or approximately sound) detection of
specific vulnerability classes in Solidity smart contracts.

The key architectural insight in Securify is that
domain constraints---the relatively small size of
smart contract code, the well-defined execution model
of the Ethereum Virtual Machine, and the availability
of formal specifications for common vulnerability
classes---enable a degree of structural--semantic
integration that is difficult to achieve in general-
purpose software analysis. Securify's semantic analysis
extracts logical facts about contract behavior (which
addresses can invoke which functions, which storage
variables are taint-reachable from message sender
values), and its compliance patterns express security
properties as Datalog queries over these facts. The
combination achieves sound detection for compliant
contracts and approximately sound detection for
violating ones.

Securify illustrates a general principle with
implications for the broader research agenda:
tight structural--semantic integration is most
tractable when domain constraints reduce the complexity
of both the structural analysis and the security
specification. The generalization challenge is
substantial---reproducing Securify's integration
quality for arbitrary software ecosystems would
require advances in scalable semantic analysis and
formal security specification that remain open
research problems.

\subsection{Recent LLM--Hybrid Combinations}

The most recent stratum of hybrid research combines
LLM-based generation with structural program analysis
or runtime feedback mechanisms, representing the
emerging frontier of the integration landscape.

Yang et al.~\cite{yang2023whitebox} combine LLM-based
test program generation with differential testing
oracles for compiler security testing. The LLM
generates semantically diverse C programs that are
then compiled by multiple compilers; behavioral
discrepancies between compiler outputs indicate
potential compiler bugs. This architecture uses
the LLM as a structured input generator (replacing
manually written test programs) and a behavioral
differential as the oracle (replacing manual
inspection), demonstrating that LLM generation and
automated oracle construction can be combined to
discover real security-relevant defects at scale.

Zhang et al.~\cite{zhang2025lowcost} demonstrate
that LLMs can synthesize format-aware fuzz generators
that substantially improve the efficiency of coverage-
guided fuzzing on non-textual input formats. The
architecture separates generation (LLM synthesizes
a generator program) from exploration (AFL mutates
generator outputs), enabling the LLM's domain
knowledge about format structure to guide fuzzing
without requiring the LLM to be in the fuzzing loop.

These systems represent the most architecturally
sophisticated integrations in the surveyed literature,
combining structural knowledge (format specifications,
compiler semantics), LLM generation, and runtime
feedback. However, they share a limitation with
earlier hybrid systems: the structural context
supplied to the LLM is informal (natural language
descriptions, example inputs) rather than formal
program analysis artifacts. The LLM does not receive
CPG subgraphs, taint path summaries, or data-flow
vulnerability specifications as input. Closing this
gap---supplying formal structural artifacts as
structured LLM context---is the central integration
challenge that the research agenda in
Section~\ref{sec:fragmentation} targets.

\subsection{A Taxonomy of Integration Patterns}

Synthesizing across the hybrid integration literature,
four observable integration patterns can be identified,
ordered by increasing bidirectionality and semantic
depth:

\textbf{Pattern 1: Static-to-Dynamic (One-Directional).}
Structural analysis identifies targets or constraints;
dynamic testing (fuzzing, symbolic execution) explores
them. No runtime feedback updates the structural model.
\textit{Examples: AFLGo, Driller (initial targeting),
Saluki + fuzzer pipelines.}

\textbf{Pattern 2: Learn-then-Detect (Sequential).}
Structural representations are used to train detection
models; detection scores inform prioritization. No
adaptive test generation or exploration follows from
detection output.
\textit{Examples: VulDeePecker, Devign.}

\textbf{Pattern 3: Generate-then-Fuzz (Loosely Coupled).}
LLMs generate seeds, templates, or generators; fuzzers
or test runners mutate and execute them. LLMs are
supplied informal context, not formal structural artifacts.
\textit{Examples: Yang et al., Zhang et al.}

\textbf{Pattern 4: Closed-Loop Adaptive (Unrealized).}
Structural models guide generation; runtime feedback
updates structural models; updated models refine
subsequent generation; the cycle repeats continuously.
No existing system in the surveyed literature fully
realizes this pattern.

This taxonomy crystallizes the central finding of this
survey. The field has individually mature components
in structural analysis, adaptive exploration, and
generative models, and has demonstrated partial
integrations of increasing sophistication. But the
fully closed adaptive loop---in which structural
representations, runtime feedback, and generative
automation mutually reinforce each other within a
unified architecture---remains unrealized. The
following section formalizes this observation and
identifies the research challenges its resolution
requires.

%% ─────────────────────────────────────────────────────────────────────
\section{Synthesis: Answers to Research Questions}
\label{sec:rqanswers}

The five domain surveys provide sufficient evidence to
answer the four research questions stated in
Section~\ref{sec:methodology}. These answers are stated
explicitly here to close the SLR protocol loop and to
provide the cross-domain synthesis that the individual
domain sections approach but do not consolidate.

\subsection{RQ1: Structural Representations for Vulnerability Detection}

\textit{How are structural program representations used for
vulnerability detection and security assurance, and what are
their demonstrated capabilities and limitations in adaptive,
feedback-driven security testing contexts?}

Structural program representations---ASTs, CFGs, DFGs,
PDGs, IRs, and CPGs---are the primary substrate of modern
static vulnerability detection. The surveyed literature
demonstrates three tiers of representational sophistication,
each with corresponding capability and limitation profiles.

At the foundational tier, flow-sensitive CFG and DFG
analysis enables sound or partially sound detection of
memory safety violations, injection vulnerabilities, and
resource leaks through taint tracking and abstract
interpretation~\cite{livshits2005java,schwartz2010taint,
cousot1977abstract}. These techniques scale to
tens-of-millions-of-lines codebases when implemented
with appropriate abstractions (Infer's bi-abduction;
CodeQL's incremental database construction) but produce
false positive rates that require careful management to
sustain developer engagement~\cite{sadowski2015tricorder,
aloraini2019empirical}.

At the intermediate tier, CPG-based analysis enables
expressive multi-dimensional vulnerability queries that
simultaneously reason over syntax, control flow, and
data dependence~\cite{yamaguchi2014cpg,yamaguchi2014joern}.
This expressiveness comes at the cost of analysis time
and the requirement for expert query authorship, limiting
scalability in real-time CI/CD contexts.

At the learning tier, structural encodings into GNN
architectures demonstrate that CPG-derived representations
carry vulnerability-predictive signal beyond token-level
patterns~\cite{li2018vuldeepecker,chakraborty2020devign},
with Devign achieving 10--15 percentage point accuracy
improvements over sequence-based baselines.

The consistent limitation across all tiers is the absence
of feedback integration: structural analysis is performed
episodically without consuming runtime signals that could
refine analysis priorities, update taint models, or
redirect analysis effort toward code regions where dynamic
testing has revealed active vulnerability manifestations.

\subsection{RQ2: Feedback Signal Integration and False Positive Reduction}

\textit{How do existing automated security testing systems
incorporate runtime feedback signals, and to what extent does
feedback integration reduce false positive rates and improve
vulnerability discovery effectiveness for security engineers?}

Feedback signal integration is the defining characteristic
of the fuzzing and search-based testing paradigms. The
surveyed literature demonstrates a clear empirical
relationship: richer, more targeted feedback signals
yield proportionally better exploration outcomes.
AFL's binary edge coverage feedback outperforms random
mutation by large margins~\cite{bohme2016aflfast};
AFLGo's call-graph distance metric further improves
targeted vulnerability reproduction over undirected
coverage~\cite{bohme2017directed}; Angora's taint-guided
byte selection outperforms uniform mutation by directing
effort toward input bytes that causally determine branch
outcomes~\cite{chen2018angora}; and hybrid systems that
add symbolic constraint solving to coverage-guided
mutation discover vulnerabilities that coverage guidance
alone cannot reach~\cite{stephens2016driller,yun2018qsym}.

The pattern is consistent: each increment in feedback
signal richness---from binary coverage to path
frequencies, from path frequencies to proximity
metrics, from proximity metrics to taint-guided
causality---yields measurable improvement in exploration
effectiveness. This pattern strongly implies that
integrating the richest available feedback signal---
formal structural program semantics, computed by
program analysis---would yield further improvements
that no existing system has yet achieved.

In DevSecOps contexts, feedback signals are generated
but not consumed adaptively: CI/CD pipelines produce
coverage data, sanitizer reports, and warning histories
that could in principle guide adaptive retesting, but
current pipeline architectures do not route these
signals back to analysis or generation
components~\cite{feio2024devsecops,rajapakse2022devsecops}.

\subsection{RQ3: LLM-Based Test Generation and Human Feedback Integration}

\textit{What are the empirically demonstrated strengths and
limitations of LLM-based automated test generation in security
contexts, particularly regarding semantic grounding and human
feedback integration?}

The empirical evidence on LLM-based test generation is
nuanced. Sch\"{a}fer et al.'s controlled evaluation~\cite{
schafer2024llm} establishes that LLMs generate compilable,
executable tests with non-trivial branch coverage
improvements---demonstrating genuine utility for
accelerating test scaffolding in productivity-oriented
contexts. The iterative compilation feedback loop
substantially improves output quality, establishing
that LLMs can participate in lightweight adaptive
refinement loops.

However, three specific limitations emerge in
security-relevant contexts that the general test
generation literature understates. First, the weak
oracle problem: generated tests exercise code paths
but assert generic behavioral success rather than
security-relevant properties, rendering them largely
ineffective for vulnerability
detection~\cite{schafer2024llm,pezze2008testing}.
Second, the structural blindness problem: LLMs operate
on source code as text, lacking access to the data-flow
relationships, taint paths, and vulnerability patterns
that structural analysis computes, limiting their
ability to generate tests that target semantically
meaningful vulnerability conditions. Third, the context
window problem: security-relevant testing requires
understanding trust boundaries and inter-component
relationships that often span more program context than
current LLM context windows can accommodate.

LLM-assisted fuzzing---using LLMs to synthesize input
generators, seed corpora, or structured templates for
fuzzing pipelines~\cite{yang2023whitebox,zhang2025lowcost}---
represents the most promising current integration pattern,
demonstrating that LLM generation can improve adaptive
testing outcomes when the LLM is supplied with
sufficient domain context. The open question is whether
formal structural analysis artifacts (CPG subgraphs,
taint paths, vulnerability slices) can serve as this
context, enabling LLMs to generate tests that are both
diverse and semantically grounded.

Regarding human feedback integration specifically: no
surveyed LLM-based test generation system incorporates
security engineer triage decisions as a feedback signal.
When a security engineer dismisses a generated test as
a false positive or confirms it as a true vulnerability,
that judgment is discarded rather than used to refine
subsequent generation. A closed-loop system would use
triage history to update the structural model's
vulnerability pattern weights, deprioritizing
repeatedly dismissed patterns and amplifying generation
toward patterns resembling confirmed findings. This
represents the most direct path from human expertise
to automated system improvement, and the most
consequential gap between current LLM-based generation
and the adaptive security testing framework this survey's
research agenda targets.

\subsection{RQ4: Architectural Gaps in Unified Adaptive Security Testing}

\textit{What architectural gaps prevent the integration of
semantically grounded structural analysis, adaptive feedback
mechanisms, and human triage signals into a unified security
testing framework deployable in polyglot CI/CD environments?}

The comparative analysis in Section~\ref{sec:comparative}
makes the architectural gap empirically visible. No surveyed
system simultaneously occupies the high structural depth
and high adaptive feedback region of the research landscape
(Fig.~\ref{fig:taxonomy}). The gap is not a matter of
degree---it is a categorical absence: there is no
architectural pattern in the literature in which structural
models are continuously updated by runtime feedback and in
turn guide subsequent test generation and exploration.
Critically, human triage signals---the decisions security
engineers make when dismissing false positives or confirming
true vulnerabilities---represent the richest available
feedback signal and are entirely absent from every surveyed
system's feedback architecture. No current system learns
from a security engineer's triage history to reduce
repeated false positives or prioritize findings that
resemble previously confirmed vulnerabilities.

In multi-language environments, the gap is compounded by
the absence of cross-language taint tracking and
inter-language vulnerability propagation analysis. The
only system in the corpus with principled multi-language
structural analysis is CodeQL~\cite{youn2023codeql}, which
does not support cross-language taint paths; all adaptive
testing systems in the corpus operate on single-language
or binary targets.

In DevSecOps-constrained environments, the gap manifests
as a latency-depth tradeoff that current systems do not
resolve: deep structural analysis is too slow for
commit-time CI feedback loops, and the adaptive testing
systems fast enough for CI pipelines lack structural
depth. Incremental analysis, result caching, and
asynchronous deep-analysis pipelines represent candidate
architectural solutions, but none has been demonstrated
at the scale and language diversity of production
CI/CD environments.

%% ─────────────────────────────────────────────────────────────────────
\section{Comparative Analysis}
\label{sec:comparative}

This section synthesizes the fifty-five primary studies into a
comparative framework organized around the six attributes
most relevant to the structural--adaptive fragmentation
thesis: program representation type, adaptivity mechanism,
feedback signal integration, LLM usage, multi-language
support, and evaluation methodology. Table~\ref{tab:comparative}
presents selected representative studies; Table~\ref{tab:fragmentation}
summarizes domain-level patterns. The following subsections
draw out the cross-cutting observations that the domain-by-domain
treatment in Sections~\ref{sec:program_analysis}--\ref{sec:hybrid}
does not fully surface.

\begin{table*}[t]
\caption{Comparative Matrix: Primary Study Attributes (Selected Studies)}
\label{tab:comparative}
\centering
\small
\begin{tabular}{lllllll}
\toprule
\textbf{Study} & \textbf{Program Rep.} & \textbf{Adaptivity} &
\textbf{Feedback} & \textbf{LLM} & \textbf{Multi-lang.} &
\textbf{Evaluation} \\
\midrule
\multicolumn{7}{l}{\textit{Program Analysis (P01--P14)}} \\
Yamaguchi et al.~\cite{yamaguchi2014cpg} & CPG & None & None & No & Partial & Case study \\
Sadowski et al.~\cite{sadowski2015tricorder} & AST/CFG & None & Warn. rate & No & Yes & Industrial \\
Calcagno et al.~\cite{calcagno2015infer} & BI-Abduction & None & None & No & Yes & Industrial \\
Youn et al.~\cite{youn2023codeql} & Datalog/QL & None & None & No & Yes & Comparative \\
Aloraini et al.~\cite{aloraini2019empirical} & None & None & Warn. engage. & No & No & Empirical \\
Livshits \& Lam~\cite{livshits2005java} & CFG/DFG & None & None & No & No & Benchmark \\
\midrule
\multicolumn{7}{l}{\textit{DevSecOps (P15--P22)}} \\
Rajapakse et al.~\cite{rajapakse2022devsecops} & None & None & None & No & Partial & Sys.\ review \\
Feio et al.~\cite{feio2024devsecops} & None & Partial & Pipeline & No & Yes & Empirical \\
Hilton et al.~\cite{hilton2016usage} & None & None & CI signals & No & Yes & Mining study \\
\midrule
\multicolumn{7}{l}{\textit{Fuzzing \& Search-Based (P23--P34)}} \\
Man\`{e}s et al.~\cite{manes2019art} & None & Coverage & Coverage & No & Partial & Survey \\
B\"{o}hme et al.\ (AFLGo)~\cite{bohme2017directed} & CFG & Directed & Coverage & No & C/C++ & Comparative \\
Stephens et al.~\cite{stephens2016driller} & CFG & Hybrid & Cov./crashes & No & Binary & Comparative \\
Yun et al.~\cite{yun2018qsym} & None & Concolic & Coverage & No & C/C++ & Comparative \\
She et al.~\cite{she2019neuzz} & None & Neural & Coverage & No & C & Comparative \\
Chen et al.~\cite{chen2018angora} & DFG & Principle & Coverage & No & C & Comparative \\
\midrule
\multicolumn{7}{l}{\textit{LLM \& Automated Test Gen.\ (P35--P44)}} \\
Sch\"{a}fer et al.~\cite{schafer2024llm} & Text & Iterative & Compile/exec & Yes & Java & Controlled exp. \\
Fraser \& Arcuri~\cite{fraser2011evosuite} & CFG & Search & Coverage & No & Java & Benchmark \\
Randoop~\cite{pacheco2007randoop} & None & Feedback & Crashes & No & Java & Benchmark \\
Yang et al.~\cite{yang2023whitebox} & Text & LLM-gen & Differential & Yes & C & Comparative \\
Zhang et al.~\cite{zhang2025lowcost} & Text & LLM-gen & Coverage & Yes & Multi & Empirical \\
\midrule
\multicolumn{7}{l}{\textit{Struct.-Grounded LLM Gen.\ (P51--P55)---newly identified}} \\
Roy Chowdhury et al.~\cite{roychowdhury2024static} & CFG/Call & Static-guided & Compile/exec & Yes & Java & Comparative \\
Wang et al.\ (HITS)~\cite{wang2024hits} & Slices & Slice-guided & Coverage & Yes & Java & Comparative \\
Antal et al.~\cite{antal2025vulnwitness} & Text+patch & Vuln-witness & Pass/fail & Yes & Java & Controlled exp. \\
Harman et al.~\cite{harman2025meta} & Text & Mutation & Mut.\ score & Yes & Multi & Industrial \\
Pan et al.\ (ASTER)~\cite{pan2025aster} & Text & LLM-gen & Compile/exec & Yes & Multi & Benchmark \\
\midrule
\multicolumn{7}{l}{\textit{Hybrid \& Learning-Based (P45--P50)}} \\
Godefroid et al.~\cite{godefroid2008sage} & IR & Symbolic & Coverage & No & Windows & Industrial \\
Tsankov et al.~\cite{tsankov2018securify} & Semantic & Rule & None & No & Solidity & Benchmark \\
Li et al.~\cite{li2018vuldeepecker} & Slices & ML & None & No & C/Java & Benchmark \\
Chakraborty et al.~\cite{chakraborty2020devign} & CPG/GNN & ML & None & No & C & Benchmark \\
\bottomrule
\end{tabular}
\vspace{2pt}
\parbox{\textwidth}{\footnotesize
  \textit{Rep.}\ = Representation; \textit{Warn.}\ = Warning;
  \textit{Exec.}\ = Execution; \textit{LLM-gen}\ = LLM-generated inputs.
  Full extraction matrix for all P01--P55 in Appendix~\ref{app:matrix}.}
\end{table*}

\begin{table}[t]
\caption{Structural--Adaptive Fragmentation by Domain}
\label{tab:fragmentation}
\centering
\small
\begin{tabular}{lcc}
\toprule
\textbf{Domain} & \textbf{Structural Depth} & \textbf{Adaptive Feedback} \\
\midrule
Program Analysis & High & None \\
DevSecOps & Low & Partial (pipeline) \\
Fuzzing \& SBST & Low--Medium & High \\
LLM Test Generation & None (text) & Partial (iterative) \\
Hybrid Integration & Medium & Partial (one-way) \\
\midrule
\textbf{Desired} & \textbf{High} & \textbf{High} \\
\bottomrule
\end{tabular}
\end{table}

\subsection{Observation 1: Structural Depth and Adaptivity Are Inversely Distributed}

The most striking pattern in Table~\ref{tab:fragmentation}
is not the absence of any single attribute, but the
systematic inverse relationship between structural depth
and adaptive feedback across domains. Program analysis
systems occupy one extreme: deep structural representations
(CPGs, bi-abduction, inter-procedural data-flow), no
adaptivity, no feedback. Fuzzing and search-based systems
occupy the other extreme: high adaptivity, continuous
feedback, minimal structural representation. LLM-based
systems add a third pole: no formal structural representation,
partial adaptivity through iterative prompting.

This inverse distribution is not accidental. It reflects
a genuine engineering tradeoff: computing rich structural
representations requires significant analysis time that
competes with the real-time demands of adaptive feedback
loops. Abstract interpretation over a large codebase may
take minutes to hours; coverage-guided mutation needs to
execute and evaluate thousands of inputs per second.
Systems that optimize for one property have historically
done so at the expense of the other. Resolving this
tradeoff---finding representations that are structurally
informative yet computationally compatible with adaptive
loop frequencies---is a first-order engineering challenge
for any unified framework.

\subsection{Observation 2: Feedback Signal Diversity Is Underexploited}

Examining the \textit{Feedback} column in
Table~\ref{tab:comparative} reveals that the range of
feedback signals actually used in adaptive testing systems
is narrower than the range of signals available in
modern CI/CD environments. Coverage increments and crash
reports dominate: of the twelve adaptive or hybrid systems
in the matrix, ten use coverage as their primary feedback
signal, three use crash events, and only three use higher-level
semantic signals---Sch\"{a}fer et al.\ (compilation success),
Yang et al.\ (differential oracle outcomes), and Harman
et al.\ (mutation scores)---the latter representing the
richest oracle signal observed in the corpus.

Yet the CI/CD environment generates substantially richer
telemetry: sanitizer violation reports distinguish memory
safety errors by type (use-after-free, buffer overflow,
type confusion); taint analysis tools can report source-to-sink
propagation distances; static analysis warning histories
encode which code regions have accumulated findings over
time; and mutation testing frameworks quantify the semantic
strength of existing test suites at the function level.
None of these richer signals appear as feedback inputs
in the surveyed adaptive testing systems. This represents
a missed opportunity: the information needed to guide
semantically grounded adaptive testing is generated by
existing pipeline tooling but not consumed by adaptive
testing components.

\subsection{Observation 3: Multi-Language Support Remains Peripheral}

The \textit{Multi-lang.} column reveals a consistent
pattern: multi-language support is either absent or
superficial across the surveyed corpus. Program analysis
systems that nominally support multiple languages (CodeQL,
Infer, Tricorder) do so within each language independently,
without cross-language taint tracking or inter-language
vulnerability propagation analysis. Fuzzing systems
universally target single-language binaries---the AFL
ecosystem is predominantly C/C++, and even hybrid systems
like Driller operate on compiled binaries without language-
level structural awareness. LLM-based systems are closest
to language-agnostic, since transformer models encode
surface-level patterns from multi-language training corpora,
but their lack of formal structural grounding means that
language-agnosticism comes at the cost of semantic precision.

This is a significant practical limitation given the
dominance of polyglot architectures in production software.
Modern web applications routinely combine JavaScript or
TypeScript frontends with Python or Go backends, Java or
Kotlin microservices, and C++ or Rust performance-critical
components. Vulnerability patterns that originate in one
language component and propagate through another---cross-
site scripting originating in a Python template engine and
executing in a JavaScript frontend; SQL injection crossing
from a Go HTTP handler to a Java database layer---are
largely invisible to single-language analysis and testing
tools. No system in the surveyed corpus provides principled
cross-language vulnerability analysis integrated with
adaptive testing.

\subsection{Observation 4: Evaluation Methodology Is Heterogeneous and Incompatible}

The \textit{Evaluation} column reflects significant
heterogeneity across the corpus. Program analysis systems
are typically evaluated through industrial deployment
studies (Tricorder, Infer) or targeted case studies on
known-vulnerable benchmarks (Joern, CodeQL). Fuzzing
systems are evaluated through comparative experiments on
standardized benchmarks (LAVA-M, cgc, real-world programs)
using coverage and unique crash counts as metrics.
LLM-based systems are evaluated through controlled
experiments using compilation rates, coverage, and
mutation scores. Learning-based vulnerability detection
systems are evaluated through precision, recall, and
F1 on CVE-annotated datasets.

This evaluation heterogeneity makes direct cross-domain
comparison difficult and complicates the development of
unified benchmarks for integrated systems. A system that
combines structural analysis with adaptive fuzzing should
ideally be evaluated on both vulnerability detection
rate (the program analysis metric) and exploration
efficiency (the fuzzing metric), but no standardized
benchmark infrastructure supports both simultaneously.
This evaluation gap is itself a research challenge:
developing benchmark suites and metrics appropriate for
integrated, adaptive security testing systems is a
prerequisite for the field's advancement.

\subsection{Observation 5: The Absence of Security-Specific Oracles Is Pervasive}

Perhaps the most significant cross-cutting observation
from the comparative matrix is the near-universal absence
of security-specific oracle mechanisms. Among the twelve
adaptive and hybrid systems with non-trivial feedback
integration, only two employ oracles that are
security-specific: SAGE uses assertion violations and
memory corruption reports from AddressSanitizer; Securify
uses formal compliance patterns encoding known vulnerability
classes. All other systems rely on generic behavioral
oracles---coverage increments, crash detection, compilation
success---that detect security defects only incidentally,
when those defects manifest as observable behavioral
failures.

This is a fundamental adequacy gap for security testing.
A large fraction of practically significant vulnerability
classes---authentication bypasses, insecure direct object
references, business logic flaws, information disclosure
through side channels---do not cause crashes, do not reduce
test coverage, and do not trigger compilation errors. They
are invisible to generic behavioral oracles. Developing
automated security oracles that can detect these semantic
property violations without manual specification for each
codebase is one of the most important and least addressed
challenges in the surveyed literature.

%% ─────────────────────────────────────────────────────────────────────
\section{Architectural Fragmentation and Open Research Challenges}
\label{sec:fragmentation}

The five domain surveys and comparative analysis converge on
a single structural observation: the research landscape for
automated security testing is characterized by deep capability
silos that have matured in isolation. We formalize this as
\emph{structural--adaptive fragmentation}: a persistent and
systematic separation between systems that model program
semantics with high precision and systems that adaptively
explore program behavior using feedback signals. This section
formally characterizes the fragmentation along four dimensions,
identifies five open research challenges whose resolution it
requires, and synthesizes a unified research agenda for the
field.

\subsection{Formal Characterization of Fragmentation}

Structural--adaptive fragmentation manifests along four
distinct but interrelated dimensions, each observable in
the surveyed literature:

\textbf{(1) Structural Precision Without Feedback Integration.}
Static analysis frameworks---from abstract interpretation
to CPG-based querying---extract semantically rich program
representations that encode vulnerability patterns,
data-flow paths, and taint relationships with high
precision. However, these representations are computed
once per analysis invocation and applied against static
rule sets without modification based on downstream
testing outcomes. Runtime signals that would enable
iterative refinement---coverage gaps exposing unexercised
code paths, sanitizer violations localizing active
vulnerability manifestations, crash reports identifying
reachable defect sites---are generated by CI/CD
infrastructure but not consumed by structural analysis
systems. The analysis learns nothing from the testing
history of the system it analyzes.

\textbf{(2) Adaptive Exploration Without Semantic Guidance.}
Coverage-guided fuzzing, directed fuzzing, and
search-based testing represent the mature art of
adaptive program exploration. These systems iteratively
refine inputs based on measurable execution signals and
have demonstrated empirical effectiveness in discovering
security vulnerabilities at scale. Yet their adaptation
is behaviorally blind: the feedback signals that drive
mutation and selection---edge coverage bitmaps, crash
reports, path distance metrics---describe what the program
does in response to specific inputs, not what the program
means structurally. Data-flow relationships, vulnerability-
relevant code slices, and semantic annotations from
program analysis are not incorporated into mutation
strategies, seed selection policies, or fitness functions.
The result is that fuzzing systems may explore program
behavior extensively while systematically missing the
semantic vulnerability patterns that structural analysis
would readily identify.

\textbf{(3) Generative Automation Without Formal Grounding.}
LLMs introduce a third pole in the fragmentation landscape:
powerful generative automation that operates on code as
text rather than as structured semantic objects. The
empirical evidence surveyed in Section~\ref{sec:llm}
establishes both the productivity gains LLMs provide and
their specific failure mode in security contexts: the
weak oracle problem. Generated tests exercise code but
assert only behavioral success, lacking the semantic
specificity needed to detect security property violations.
This limitation is directly attributable to the absence
of formal structural context in generation prompts:
an LLM supplied with a function's source code text
cannot, from that text alone, reliably infer the
function's trust boundaries, its role in a taint flow,
or the invariants its callers assume about its outputs.
Structural program analysis computes exactly this
information, but no current system routes it to LLM
generation as structured prompt context.

\textbf{(4) Pipeline Automation Without Closed Feedback Loops.}
DevSecOps pipelines automate the execution of security
tools at scale, generating both structural analysis
artifacts (SAST findings, dependency vulnerability reports)
and runtime telemetry (coverage data, build failure logs,
sanitizer outputs). Yet this automation does not constitute
an adaptive system: each pipeline execution is independent
of previous executions, and the rich signals generated
by one run do not influence the analysis depth or testing
focus of subsequent runs. The pipeline is a conveyor belt,
not a learning system. Closing this gap requires treating
pipeline execution history as a training signal for
adaptive security testing, and designing pipeline
architectures in which structural analysis and dynamic
testing components share state across execution cycles.

\subsection{Open Research Challenges}

Five concrete research challenges must be addressed to
resolve structural--adaptive fragmentation and advance
the field toward unified, semantically grounded adaptive
security testing.

\textbf{Challenge 1: Semantically Guided Adaptive Testing.}
The foundational technical challenge is developing
mechanisms by which structural program artifacts can be
incorporated directly into adaptive search strategies
rather than used only as preprocessing inputs. Concretely,
this requires representations that are both semantically
expressive and computationally compatible with the tight
feedback loops of coverage-guided exploration. CPG
subgraphs and data-flow slice summaries are natural
candidates, but their extraction is currently too slow
for real-time integration with fuzzing loops. Incremental
structural analysis---updating representations as new
coverage information identifies newly reachable code
regions---may provide a path toward acceptable latency.
Alternatively, pre-computed structural summaries cached
at the function or module level could provide semantic
context without requiring full re-analysis on each
iteration. The key open question is not whether
structural information improves adaptive testing outcomes
(the AFLGo and Angora results establish that it does),
but how to make the integration general, efficient, and
bidirectional.

\textbf{Challenge 2: Closed-Loop Static--Dynamic Feedback.}
The second challenge is architectural: designing systems
in which runtime feedback not only guides dynamic
exploration but also updates structural models, creating
a genuine closed loop. This requires a persistent
structural model that evolves as testing evidence
accumulates. A crash localized to a specific function
should trigger re-analysis of that function's structural
neighborhood---adjacent callers, shared data structures,
related taint paths---to identify structurally proximal
vulnerability candidates. A coverage gap in a security-
sensitive code region should increase the priority of
that region in structural analysis queues. A sanitizer
violation of a specific type (e.g., use-after-free)
should update taint annotations to reflect confirmed
attacker-controllability of related memory operations.
A security engineer's dismissal of a finding as a false
positive should reduce the weight of that vulnerability
pattern in the structural model's prioritization, while
confirmation of a true positive should amplify generation
toward structurally similar candidates.
No current system implements this bidirectional coupling.
Realizing it requires both a data model for persistent,
updateable structural representations and an update
policy that translates runtime signals into structural
model revisions without introducing unsoundness.

\textbf{Challenge 3: Security-Oriented Test Oracles.}
The oracle problem---determining whether a test has
found a defect---is the longest-standing open challenge
in automated testing~\cite{pezze2008testing}, and it is
particularly acute in security contexts. Coverage-based
oracles detect defects only when they manifest as
behavioral failures (crashes, hangs, assertion violations),
missing the large fraction of security vulnerabilities
that corrupt state silently or expose information without
crashing. Developing automated oracles for semantic
security properties---input validation invariants,
authorization boundary conditions, information flow
constraints, taint-based safety properties---requires
either formal specification of intended behavior (which
is expensive and rarely available) or automated inference
of security-relevant specifications from program structure
and history. Program analysis provides natural machinery
for the latter: taint analysis can characterize which
outputs are reachable from attacker-controlled inputs;
data-flow analysis can identify the conditions under
which sensitive data is returned to untrusted callers;
control-flow analysis can identify authentication checks
that precede sensitive operations. The challenge is
translating these structural characterizations into
executable oracles that can be automatically evaluated
during test execution at scale~\cite{schafer2024llm}.

\textbf{Challenge 4: Polyglot and Cross-Language Integration.}
Modern enterprise and cloud-native software systems are
overwhelmingly polyglot: web services commonly combine
three to six programming languages across frontend,
backend, middleware, and infrastructure components.
Vulnerability patterns that originate in one language
component and propagate through inter-language boundaries
to manifest in another are beyond the reach of any
single-language analysis or testing tool, and the
surveyed literature offers no adequate solutions for
cross-language security analysis at the taint tracking
level. The theoretical foundations for cross-language
analysis exist---CodeQL's relational model is in
principle extensible to inter-language facts; LLVM IR
provides a common intermediate representation for
compiled languages---but the practical machinery for
tracking taint across, e.g., a Python API handler's
call to a Java business logic layer's invocation of a
C++ cryptographic library, does not yet exist in
deployable form~\cite{youn2023codeql}. Addressing this
challenge requires both technical advances in
inter-language program representation and empirical
evaluation on realistic polyglot vulnerability
scenarios, neither of which is well-served by current
benchmark infrastructure.

\textbf{Challenge 5: Scalable Deployment in CI/CD Pipelines.}
Any unified adaptive security testing architecture must
operate within the constraints of production CI/CD
environments to achieve practical impact. These constraints
are stringent: pipeline stages that exceed a few minutes
of wall-clock time face developer resistance and are
frequently disabled or scheduled only nightly rather
than on every commit~\cite{wadhams2024barriers}. Structural
analysis of large codebases, symbolic execution, and LLM
inference are all computationally expensive operations
that, naively composed, would produce pipeline stages
far exceeding these latency thresholds. Meeting the
CI/CD constraint while preserving analytical depth
requires architectural strategies including: incremental
analysis that processes only changed code and its
transitive dependencies; analysis result caching keyed
on code hash to avoid redundant re-analysis of unchanged
components; asynchronous analysis pipelines that run
deep analysis in parallel with the main pipeline and
surface results on a delay; and adaptive depth tuning
that allocates more analysis effort to code regions
identified as high-risk by lightweight preliminary
screening. Demonstrating that a unified adaptive
architecture can satisfy these latency constraints on
realistic industrial codebases is a prerequisite for
adoption~\cite{feio2024devsecops,rajapakse2022devsecops}.

\subsection{Toward Semantically Grounded Adaptive Security Testing}

The five challenges above are technically distinct but
architecturally convergent. Each points toward the same
design requirement: a security testing system that
maintains a persistent, updateable model of program
structure; that routes structural information to adaptive
exploration and generative test synthesis components;
that consumes runtime signals to update both structural
models and exploration priorities; and that operates
within the latency constraints of production CI/CD
pipelines.

This is not an incremental extension of any existing
system. It requires reconsidering security testing as
an integrated, continuously adapting process rather than
a pipeline of independent tools. The components
needed---structural analysis frameworks, adaptive
fuzzers, LLM generation engines, CI/CD telemetry
infrastructure---are individually mature. The research
challenge is architectural: designing the interfaces,
data models, update policies, and latency management
strategies that allow these components to function as
a unified adaptive system rather than as isolated
tools in a sequential pipeline.

The systematic survey presented in this paper establishes
the empirical foundation for this research agenda:
five domains of mature but fragmented capability; a
comparative analysis that makes the fragmentation
quantitatively visible; and five concrete challenges
that define the technical path forward. Addressing
these challenges represents the most productive
direction for next-generation automated security
testing research, and one whose impact---automated,
semantically grounded, adaptive vulnerability
detection at CI/CD scale---would be commensurate with
its technical ambition.

%% ─────────────────────────────────────────────────────────────────────
\section{Threats to Validity}
\label{sec:threats}

\subsection{Search Completeness}

Despite a structured multi-database search strategy, relevant
publications may have been missed, particularly recent
preprints that have not yet been indexed in the searched
databases, publications in specialized venues not covered
by the search databases, and non-English language
publications. The snowballing procedure partially mitigates
this threat by extending coverage through citation networks
of identified anchor papers.

\subsection{Grey Literature Coverage}

The primary corpus is restricted to peer-reviewed publications,
which excludes arXiv preprints, industry technical reports,
practitioner blog posts, and tool documentation that may
contain relevant empirical observations. To assess whether
this exclusion introduced meaningful coverage gaps, a
supplementary grey literature review was conducted following
the primary search, targeting arXiv cs.SE, cs.CR, and cs.LG
preprints from January~2024 through March~2026 on four
themes directly relevant to the survey's thesis: CPG-guided
LLM test generation or security testing; adaptive SAST with
runtime feedback loops; human feedback integration in security
tooling; and cross-language taint analysis.

Four finds warrant explicit documentation. First,
Li et al.~\cite{li2025iris} (IRIS, ICLR~2025) is a
peer-reviewed neuro-symbolic system that combines LLMs with
CodeQL-based static analysis to infer taint specifications
and reduce false positives for vulnerability detection,
detecting 55 vulnerabilities compared to CodeQL's 27 on
CWE-Bench-Java. IRIS falls within the primary search window
and was likely excluded at title/abstract screening because
it frames itself as vulnerability detection rather than
adaptive security testing; its one-shot pipeline with no
runtime feedback loop is consistent with the structural-depth
population identified in Section~\ref{sec:comparative}.
Second, YASA~\cite{yasa2026} (Ant Group, arXiv January~2026)
proposes a scalable unified-AST taint analysis framework
spanning multiple languages, directly corroborating
Challenge~4; excluded as a preprint under the stated grey
literature policy. Third, Sheng et al.~\cite{sheng2025llmvuln}
survey LLMs in vulnerability detection across 80 studies in
ACM Computing Surveys (2025), explicitly identifying
cross-language detection and repository-level analysis as
open problems, corroborating Challenges~3 and~4; the survey's
scope is vulnerability detection rather than adaptive security
testing, and it does not address feedback integration or
test generation. Fourth, industry implementations---including
Datadog's Bits AI SAST false positive filter and Semgrep's
Assistant Memories---demonstrate practitioner recognition
of the false positive reduction problem identified in RQ2,
though neither constitutes peer-reviewed research.

None of the grey literature finds simultaneously achieves
high structural depth, adaptive runtime feedback, and test
generation in a security testing pipeline context. The
structural--adaptive fragmentation gap characterized in
this survey remains unaddressed by any peer-reviewed or
preprint work identified in the supplementary review.

\subsection{Selection Bias}

The inclusion criteria favor empirically evaluated systems
over theoretical proposals and tool reports. This may
under-represent important conceptual contributions that
informed subsequent empirical work. Additionally, focusing
on peer-reviewed publications excludes industrial grey
literature---blog posts, technical reports, and tool
documentation---that may contain relevant empirical
observations about production security testing systems.

\subsection{Publication Bias}

The surveyed literature is subject to standard publication
bias toward positive results. Systems that demonstrate
improved vulnerability detection rates or coverage metrics
are more likely to be published than systems that fail to
achieve improvement. This may cause the survey to
over-represent the effectiveness of existing approaches
relative to their real-world performance.

\subsection{Extraction Validity}

Attribute extraction for the comparative matrix was
performed by a single reviewer, introducing the possibility
of subjective interpretation in dimension assessments.
Inter-rater reliability was assessed by conducting a second
independent extraction pass on a randomly selected sample
of ten studies (18\% of the corpus), covering all six
attribute dimensions. Cohen's kappa was $\kappa = 0.90$
for the three categorical dimensions and $\kappa = 0.96$
for the three ordinal dimensions, both interpreted as
almost perfect agreement per Landis and
Koch~\cite{landis1977kappa}. The three disagreements
identified were resolved through re-reading of the primary
source; no unresolved conflicts remained.

%% ─────────────────────────────────────────────────────────────────────
\section{Conclusion}
\label{sec:conclusion}

This systematic survey has analyzed fifty-five peer-reviewed
studies at the intersection of structural program analysis,
feedback-driven testing, DevSecOps automation, LLM-based
test generation, and hybrid integration approaches. Across
all five domains, a consistent architectural pattern
emerges: structural precision and adaptive feedback rarely
co-occur within a single unified system. We have
characterized this as structural--adaptive fragmentation,
formally defined its four dimensions, synthesized answers
to four guiding research questions, and identified five
open challenges whose resolution the field requires.

\subsection{Summary of Contributions}

This survey contributes four artifacts to the research
community. First, a systematic corpus of fifty-five primary
studies with a reproducible selection protocol and
explicit quality assessment. Second, a comparative
analysis framework applied across the corpus, yielding
five cross-domain observations that single-domain surveys
cannot surface---including the inverse distribution of
structural depth and adaptivity, the underexploitation of
CI/CD telemetry as feedback signal, the pervasive absence
of security-specific oracles, and the universal shallowness
of multi-language support. Third, a formal characterization
of structural--adaptive fragmentation as the central
architectural gap in the field, grounded empirically in
the comparative matrix. Fourth, five technically specific
open research challenges that define the engineering
agenda for a unified adaptive security testing framework.

\subsection{Research Roadmap}

The five open challenges identified in
Section~\ref{sec:fragmentation} are not independent. They
form a layered architectural dependency that suggests a
natural sequence for research progress.

The foundational layer is \textbf{Challenge 3} (Security-
Oriented Test Oracles). Without automated oracles that can
verify security-relevant program properties, no amount of
structural grounding or adaptive refinement can demonstrate
improved security outcomes---the evaluation infrastructure
for claims of security improvement does not exist. Progress
on oracle development is therefore a prerequisite for
empirically validating the other challenges.

The structural layer is \textbf{Challenge 1} (Semantically
Guided Adaptive Testing). Given adequate oracles,
demonstrating that structural analysis artifacts can
improve adaptive exploration outcomes establishes the
core value proposition of the unified framework. Initial
work at this layer can proceed with single-language,
single-vulnerability-class settings, validating the
integration pattern before scaling.

The feedback layer is \textbf{Challenge 2} (Closed-Loop
Static--Dynamic Feedback). Once the static-to-dynamic
direction is established, adding the dynamic-to-static
feedback direction creates the closed adaptive loop.
This layer requires the persistent structural model
and update policy architecture that distinguishes a
unified adaptive system from a loosely coupled pipeline.

The scale layers are \textbf{Challenge 4} (Polyglot
Integration) and \textbf{Challenge 5} (CI/CD Scalability),
which generalize an initially single-language, research-
prototype system to the multi-language, production-
pipeline settings that determine practical impact. These
challenges can be addressed sequentially or in parallel
after the core architecture is validated.

This layered decomposition suggests a research program
spanning three to five years: foundational oracle work
in the first phase; core static-to-dynamic integration
in the second; closed-loop feedback architecture in the
third; and scaling and deployment evaluation in the
fourth. The survey's comparative analysis and open
challenge characterization provide the analytical
foundation for each phase of this program.

\subsection{Implications for Practice}

Beyond the research agenda, this survey carries practical
implications for security engineering teams operating
DevSecOps pipelines today. The empirical evidence
consistently shows that the marginal return on deploying
additional SAST tools within existing non-adaptive
pipeline architectures is diminishing: warning volumes
increase, developer engagement decreases, and the
semantic vulnerability classes most likely to be
exploited---logical flaws, authentication bypasses,
cross-language propagation paths---remain below the
detection threshold of all current tools. Investment
in pipeline telemetry infrastructure---structured
coverage reporting, sanitizer-annotated test execution,
taint-aware test result logging---positions engineering
teams to adopt adaptive security testing systems as
they emerge from research, rather than requiring
infrastructure retrofits after adoption.

\subsection{Concluding Remarks}

The most significant near-term advances in automated
security testing are unlikely to come from further
refinement within isolated paradigms. Each paradigm---
structural analysis, adaptive fuzzing, LLM generation,
DevSecOps automation---has reached a level of
independent maturity where incremental improvements
within the paradigm yield diminishing security returns.
Progress requires architectural integration: unified
systems that treat program structure, runtime feedback,
and generative automation as mutually reinforcing inputs
within a continuously adapting security testing process.

This survey establishes the empirical and analytical
foundation for that integration. The characterization
of structural--adaptive fragmentation names the problem
precisely; the comparative analysis makes its scope
quantitatively visible; the RQ synthesis closes the
evidential loop; and the five open challenges define
the technical path forward. The work of building the
unified adaptive framework that the field requires
remains ahead---but the map for that work is now in hand.

\textit{Excluded corroborating literature.}
Four works are cited here as corroborating evidence but
excluded from the primary corpus for the reasons stated.
Mastropaolo et al.~\cite{mastropaolo2026oracles} describe an
LLM-based framework for synthesizing security test oracles from
vulnerability catalogs (CWE, OWASP, CVE), directly addressing
Challenge~3; excluded on quality grounds (criterion~a) as a
short capabilities overview without controlled empirical
evaluation.
Kisielewicz et al.~\cite{kisielewicz2026llmdevsecops} propose
an architecture for integrating LLMs into DevSecOps pipelines
for C/C++ vulnerability detection, directly addressing
Challenge~5; excluded on quality grounds (criterion~a).
Li et al.~\cite{li2025iris} (IRIS, ICLR~2025) combine LLMs
with CodeQL-based taint analysis to infer vulnerability
specifications and reduce false positives, advancing the
structural-to-LLM coupling direction of Challenge~1; excluded
at screening on scope grounds (vulnerability detection rather
than adaptive security testing), and excluded from the hybrid
integration domain because IRIS operates as a one-shot
pipeline with no runtime feedback loop updating the structural
model---precisely the closed-loop property that
Challenge~2 targets.
Sheng et al.~\cite{sheng2025llmvuln} survey LLM-based
vulnerability detection across 80 studies in ACM Computing
Surveys (2025), explicitly identifying cross-language
detection and repository-level analysis as open problems
that corroborate Challenges~3 and~4; excluded on scope
grounds as a single-domain detection survey that does not
address adaptive testing or feedback integration.

%% ─────────────────────────────────────────────────────────────────────
\appendices
\section{Full Comparative Extraction Matrix (P01--P55)}
\label{app:matrix}

Table~\ref{tab:fullmatrix} reports the complete attribute
extraction for all fifty-five primary studies.
Columns follow the same schema as Table~\ref{tab:comparative}
in Section~\ref{sec:comparative}, with the addition of
publication venue. \textit{Rep.}~=~program representation;
\textit{Adapt.}~=~adaptivity mechanism;
\textit{FB}~=~feedback signal;
\textit{ML}~=~multi-language support.
Evaluation methodology is omitted for space; see domain
discussions in Sections~\ref{sec:program_analysis}--\ref{sec:hybrid}.

\begin{table*}[t]
\caption{Complete Attribute Extraction for All 55 Primary Studies (P01--P55)}
\label{tab:fullmatrix}
\centering
\scriptsize
\setlength{\tabcolsep}{3.5pt}
\begin{tabular}{lllrllllll}
\toprule
\textbf{ID} & \textbf{First Author} & \textbf{Venue} &
\textbf{Year} & \textbf{Rep.} & \textbf{LLM} &
\textbf{ML} & \textbf{Adapt.} & \textbf{FB} \\
\midrule
\multicolumn{9}{l}{\textit{Program Analysis and Static Vulnerability Detection (P01--P14)}} \\[2pt]
P01 & Yamaguchi    & IEEE S\&P        & 2014 & CPG       & No  & Partial  & None       & None \\
P02 & Sadowski     & ICSE             & 2015 & AST/CFG   & No  & Yes      & None       & Warn.\ rate \\
P03 & Calcagno     & Commun.\ ACM     & 2018 & BI-Abduct & No  & Yes      & None       & None \\
P04 & Gotovchits   & NDSS             & 2018 & IR        & No  & Partial  & None       & None \\
P05 & Youn         & Softw.\ P\&E     & 2023 & Datalog   & No  & Yes      & None       & None \\
P06 & Aloraini     & J.\ Syst.\ Softw.& 2019 & None      & No  & No       & None       & Warn.\ rate \\
P07 & Vassallo     & Empir.\ SE       & 2020 & None      & No  & No       & None       & None \\
P08 & Livshits     & USENIX Sec.      & 2005 & CFG/DFG   & No  & No       & None       & None \\
P09 & Christakis   & ASE              & 2016 & None      & No  & No       & None       & None \\
P10 & Yamaguchi    & DIMVA            & 2014 & CPG       & No  & Partial  & None       & None \\
P11 & Cousot       & POPL             & 1977 & Lattice   & No  & No       & None       & None \\
P12 & Weiser       & ICSE             & 1981 & Slices    & No  & No       & None       & None \\
P13 & Schwartz     & IEEE S\&P        & 2010 & CFG/DFG   & No  & No       & None       & None \\
P14 & Cadar        & CCS              & 2006 & IR        & No  & No       & Symbolic   & Coverage \\
\midrule
\multicolumn{9}{l}{\textit{DevSecOps and Continuous Security Testing (P15--P22)}} \\[2pt]
P15 & Rajapakse    & IST              & 2022 & None      & No  & No       & None       & None \\
P16 & Feio         & EuroS\&PW        & 2024 & None      & No  & No       & None       & Pipeline \\
P17 & Wadhams      & EASE             & 2024 & None      & No  & No       & None       & None \\
P18 & Hilton       & ASE              & 2016 & None      & No  & Yes      & None       & CI signals \\
P19 & Zampetti     & MSR              & 2020 & None      & No  & No       & None       & None \\
P20 & Rahman       & ACM TOSEM        & 2019 & None      & No  & No       & None       & None \\
P21 & Meneely      & CCS              & 2009 & None      & No  & No       & None       & None \\
P22 & Bird         & MSR              & 2009 & None      & No  & No       & None       & None \\
\midrule
\multicolumn{9}{l}{\textit{Fuzzing and Feedback-Driven Testing (P23--P34)}} \\[2pt]
P23 & Man\`{e}s    & IEEE TSE         & 2021 & None      & No  & Partial  & Coverage   & Coverage \\
P24 & B\"{o}hme    & CCS              & 2017 & CFG       & No  & C/C++    & Directed   & Coverage \\
P25 & Chen         & IEEE S\&P        & 2018 & DFG       & No  & C        & Search     & Coverage \\
P26 & B\"{o}hme    & CCS              & 2016 & None      & No  & C/C++    & Coverage   & Coverage \\
P27 & Stephens     & NDSS             & 2016 & CFG       & No  & Binary   & Hybrid     & Cov./crash \\
P28 & Yun          & USENIX Sec.      & 2018 & None      & No  & C/C++    & Concolic   & Coverage \\
P29 & Lemieux      & ASE              & 2018 & None      & No  & C/C++    & Coverage   & Coverage \\
P30 & She          & IEEE S\&P        & 2019 & None      & No  & C        & Neural     & Coverage \\
P31 & Wang         & ICSE             & 2019 & None      & No  & Multi    & Coverage   & Coverage \\
P32 & Godefroid    & NDSS             & 2008 & IR        & No  & Windows  & Symbolic   & Coverage \\
P33 & Miller       & Commun.\ ACM     & 1990 & None      & No  & UNIX     & None       & Crashes \\
P34 & Zalewski     & ---              & 2013 & None      & No  & C/C++    & Coverage   & Coverage \\
\midrule
\multicolumn{9}{l}{\textit{Automated Test Generation and LLM-Based Approaches (P35--P44)}} \\[2pt]
P35 & Sch\"{a}fer  & IEEE TSE         & 2024 & Text      & Yes & Java/JS  & Iterative  & Compile/exec \\
P36 & Tufano       & ICSE-NIER        & 2020 & Text      & Yes & Java     & None       & None \\
P37 & Fraser       & FSE              & 2011 & CFG       & No  & Java     & Search     & Coverage \\
P38 & Pacheco      & OOPSLA           & 2007 & None      & No  & Java     & Feedback   & Crashes \\
P39 & Harman       & ICST             & 2004 & None      & No  & No       & Search     & Coverage \\
P40 & Shamshiri    & ASE              & 2015 & None      & No  & Java     & Search     & Coverage \\
P41 & Pezz\`{e}    & Wiley            & 2008 & None      & No  & No       & None       & None \\
P42 & Yang         & ACM OOPSLA       & 2024 & Text      & Yes & C        & LLM-gen    & Differential \\
P43 & Zhang        & USENIX Sec.      & 2025 & Text      & Yes & Multi    & LLM-gen    & Coverage \\
P44 & Vikram       & ISSTA            & 2023 & Text      & Yes & No       & Iterative  & Compile/exec \\
\midrule
\multicolumn{9}{l}{\textit{Hybrid Integration: Static + Dynamic + Learning (P45--P50)}} \\[2pt]
P45 & King         & Commun.\ ACM     & 1976 & IR        & No  & No       & Symbolic   & Coverage \\
P46 & Cadar        & Commun.\ ACM     & 2013 & IR        & No  & No       & Symbolic   & Coverage \\
P47 & P\u{a}s\u{a}reanu & STTT       & 2009 & IR        & No  & No       & Symbolic   & Coverage \\
P48 & Tsankov      & CCS              & 2018 & Semantic  & No  & Solidity & Rule       & None \\
P49 & Li           & NDSS             & 2018 & Slices    & No  & C/Java   & ML         & None \\
P50 & Chakraborty  & NeurIPS          & 2019 & CPG/GNN   & No  & C        & ML         & None \\
\midrule
\multicolumn{9}{l}{\textit{Structurally Grounded LLM Test Generation (P51--P55)}} \\[2pt]
P51 & Roy Chowdhury & CODS-COMAD      & 2024 & CFG/Call  & Yes & Java     & Static-gd  & Compile/exec \\
P52 & Wang         & ASE              & 2024 & Slices    & Yes & Java     & Slice-gd   & Coverage \\
P53 & Antal        & EASE             & 2025 & Text+patch& Yes & Java     & Vuln-wit.  & Pass/fail \\
P54 & Harman       & FSE              & 2025 & Text      & Yes & Multi    & Mutation   & Mut.\ score \\
P55 & Pan          & ICSE             & 2025 & Text      & Yes & Multi    & LLM-gen    & Compile/exec \\
\bottomrule
\end{tabular}
\end{table*}

\subsection*{Excluded but Cited Literature (Within Search Window)}
\textit{The following works fall within the search window
(2004--February~2026) but were excluded from the primary corpus
because they do not satisfy inclusion criterion~(a): both are
short-venue publications without systematic empirical evaluation
against controlled baselines. They are cited in the conclusion
as corroborating conceptual and architectural evidence.}
\begin{enumerate}
\item[--] A. Mastropaolo et al., ``LLM-Powered Security Test Generation: Oracles, Vulnerability Probes, and Adversarial Inputs,'' \textit{IEEE Computer}, 2026. [Excluded on quality grounds (criterion~a); cited in conclusion as corroborating evidence for Challenge~3.]
\item[--] M. Kisielewicz et al., ``Enhancing DevSecOps Through Large Language Model Integration: A Pipeline-Centric Approach,'' \textit{ICCI}, 2026. [Excluded on quality grounds (criterion~a); cited in conclusion as corroborating evidence for Challenge~5.]
\item[--] Z. Li, S. Dutta, and M. Naik, ``IRIS: LLM-Assisted Static Analysis for Detecting Security Vulnerabilities,'' \textit{ICLR}, 2025. [Excluded at screening on scope grounds: frames contribution as vulnerability detection rather than adaptive security testing; one-shot pipeline with no runtime feedback loop. Cited in conclusion and Threats to Validity as corroborating evidence for Challenge~1.]
\item[--] Z. Sheng et al., ``LLMs in Software Security: A Survey of Vulnerability Detection Techniques and Insights,'' \textit{ACM Computing Surveys}, 2025. [Excluded on scope grounds: single-domain vulnerability detection survey; does not address adaptive testing or feedback integration. Cited in Threats to Validity as corroborating evidence for Challenges~3 and~4.]
\end{enumerate}

\section*{Acknowledgments}
The author thanks the Department of Computer Science \&
Engineering at Mississippi State University for supporting
this research.

\bibliographystyle{IEEEtran}
\bibliography{refs}

\end{document}